\RequirePackage{fix-cm}

\documentclass[natbib]{svjour3_src}

\smartqed                  
\usepackage{graphicx}
\usepackage{mathptmx}      
\usepackage{aps-bibstyle}  

\journalname{Earth, Moon, and Planets}

\begin{document}

\title{Predictions for Dusty Mass Loss from Asteroids during
Close Encounters with Solar Probe Plus}
\titlerunning{Predictions for Dusty Mass Loss from Asteroids}

\author{Steven R. Cranmer}
\authorrunning{S. R. Cranmer}

\institute{S. R. Cranmer \at
Department of Astrophysical and Planetary Sciences,
Laboratory for Atmospheric and Space Physics,
University of Colorado, Boulder, CO 80309, USA \\
Tel.: +1-303-735-1265 \\
\email{steven.cranmer@colorado.edu}}

\date{Received: 17 January 2016 / Accepted: 5 June 2016}

\maketitle

\begin{abstract}
The {\em{Solar Probe Plus}} ({\em{SPP}}) mission will explore the
Sun's corona and innermost solar wind starting in 2018.
The spacecraft will also come close to a number of Mercury-crossing
asteroids with perihelia less than 0.3 AU.
At small heliocentric distances, these objects may begin to lose mass,
thus becoming ``active asteroids'' with comet-like comae or tails.
This paper assembles a database of 97 known Mercury-crossing asteroids
that may be encountered by {\em SPP,} and it presents estimates of their
time-dependent visible-light fluxes and mass loss rates.
Assuming a similar efficiency of sky background subtraction as was
achieved by {\em STEREO,} we find that approximately 80\% of these
asteroids are bright enough to be observed by the
Wide-field Imager for {\em SPP} (WISPR).
A model of gas/dust mass loss from these asteroids is developed and
calibrated against existing observations.
This model is used to estimate the visible-light fluxes and
spatial extents of spherical comae.
Observable dust clouds occur only when the asteroids
approach the Sun closer than 0.2 AU.
The model predicts that during the primary {\em SPP} mission between
2018 and 2025, there should be 113 discrete events (for 24 unique asteroids)
during which the modeled comae have angular sizes resolvable by WISPR.
The largest of these correspond to asteroids 3200 Phaethon,
137924, 155140, and 289227, all with angular sizes of roughly
15 to 30 arcminutes.
We note that the {\em SPP} trajectory may still change, but no matter
the details there should still be multiple opportunities for fruitful
asteroid observations.

\end{abstract}

\section{Introduction}
\label{sec:intro}

Asteroids and comets are probes of the primordial solar system.
Their weak gravitational attraction enables the study of
a range of physical processes that are not possible to detect on
larger moons and planets.
In recent years, the traditional astronomical distinction between
rocky asteroids and ice-rich comets has been replaced by the
recognition of a continuous distribution in both composition and
volatility \citep[e.g.,][]{We89}.
Some objects that were initially identified as asteroids have been
seen to exhibit cometary outbursts \citep{Ha90,Mz11}.
On the other hand, some known
comets have become dormant as they apparently exhausted
their volatile-rich outer layers \citep{Jn08,Ye16}.
Recent work on {\em active asteroids}
\citep{Je12,Je13,Je15,Ag16}
has shown that dusty mass loss may occur even when virtually no
icy material is left.

When small solid bodies approach the Sun, there are strong radiative
and thermal effects that release gas molecules, dust particles,
and larger pieces of regolith \citep[e.g.,][]{Db14}.
At small heliocentric distances---e.g., for sungrazing
comets---the dust dissociates rapidly and the gas becomes ionized
\citep{Pv03,BP12} and often the entire object is destroyed
\citep{Bi02}.
Mass loss from comets in the inner heliosphere remains a useful,
albeit indirect, probe of the solar wind \citep{BS00,Hu07}
and the hot solar corona \citep{Ra14}.
Many similar ablative processes may also be occurring in the
environments of extrasolar planets that orbit close to their host
stars \citep{Mu11,Mt15}.

Active asteroids in the innermost heliosphere have not yet been
explored by planetary spacecraft.
However, {\em Solar Probe Plus} ({\em{SPP}}) will spend several
years inside the orbit of Venus \citep{Mc07,Fx15} with a
minimum perihelion distance of 0.0459~AU (i.e., 9.86 solar radii).
In addition to a suite of {\em in~situ} plasma and field instruments,
the the Wide-field Imager for {\em SPP}
\citep[WISPR,][]{WISPR} will observe visible-light photons over
large fields of view.
The primary goal of WISPR is to observe K-corona emission from
Thomson-scattered electrons and F-corona emission from dust,
but it will also search for sungrazing comets and putative
Vulcanoids \citep[see, e.g.,][]{St13}.

This paper explores the ability of instruments such as WISPR
to observe extended emission from mass-losing active asteroids
in the inner heliosphere.
Sect.\  \ref{sec:orb} of this paper surveys the orbital properties
of 97 Mercury-crossing asteroids, in both the Apollo ($a > 1$~AU)
and Aten ($a < 1$~AU) groups, that could be encountered by {\em SPP.}
The sizes and visible-light fluxes of these asteroids are estimated
in Sect.\  \ref{sec:props} and compared with expected background
levels of zodiacal light.
Sect.\  \ref{sec:massloss} presents a model for the mass loss
rate of high latent-heat silicate material from the selected
asteroids, and
Sect.\  \ref{sec:tail} estimates the observable spatial extent
of dusty coma/tail regions that WISPR can resolve.
Lastly, Sect.\  \ref{sec:conc} discusses some of the the broader
implications of this work and gives suggestions for
future improvements in the modeling.

\section{Orbital Analysis}
\label{sec:orb}

\begin{figure}
\hspace*{0.20in}
\includegraphics[width=0.85\textwidth]{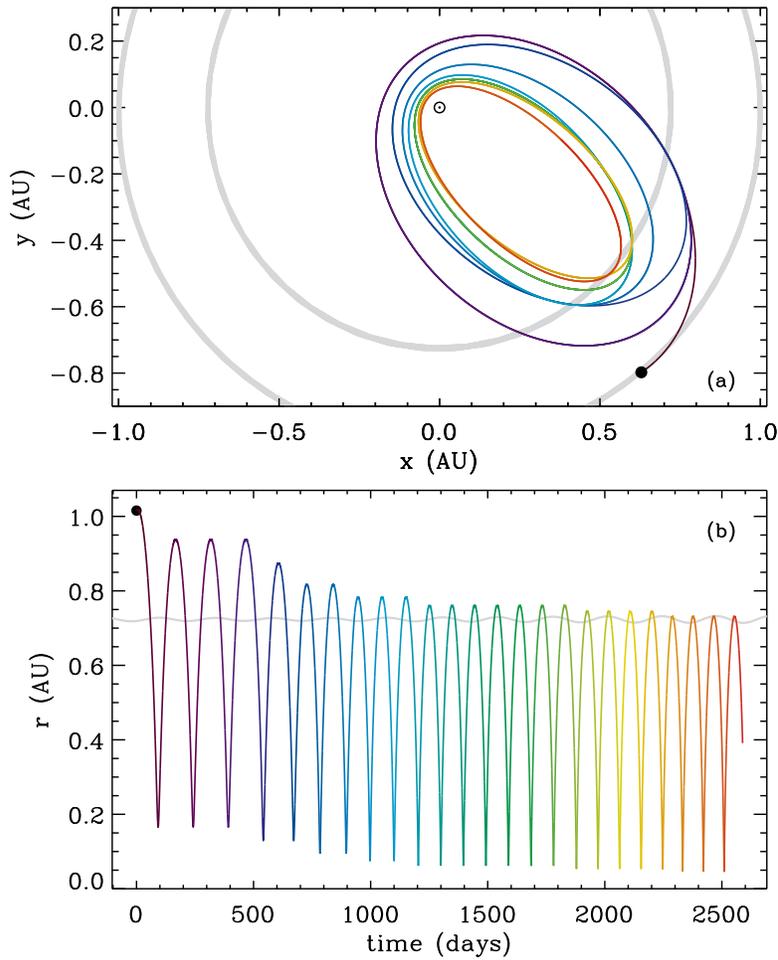}
\caption{(a) Planned {\em SPP} trajectory shown projected in
the ecliptic plane, compared with orbits of Venus and the Earth
(gray tracks).
(b) Heliocentric radial distance of {\em SPP} shown versus mission
time in days.  In both panels, the position of the Earth on
July 31, 2018 is indicated with a black circle, and mission time
is mapped to curve color.}
\label{fig01}
\end{figure}

Figure~\ref{fig01} illustrates the latest version of the proposed
{\em SPP} baseline mission trajectory.
This information was extracted from a SPICE kernel file distributed
to the {\em SPP} team in September 2014.
It assumes a launch date of July 31, 2018, and it
extends to September 1, 2025.
The kernel data were processed with version N65 of the SPICE toolkit
for IDL \citep{Ac96}, which was created by the NASA/JPL Navigation
and Ancillary Information Facility (NAIF).\footnote{%
http://naif.jpl.nasa.gov/naif/}
The Cartesian $(x,y,z)$ positions of {\em SPP}, in
solar-system barycenter coordinates for epoch J2000.0,
were computed and saved at 0.1~day intervals.
The mission time $t$ shown in Figure~\ref{fig01}(b) is specified in
days measured from 0:00 UT on July 31, 2018.

The dynamical properties of 97 Mercury-crossing asteroids were
obtained from the NASA/JPL Horizons On-Line Ephemeris System
\citep{Gg96,Gg11}\footnote{http://ssd.jpl.nasa.gov/horizons.cgi}
in July 2015.
The Horizons system was used to convert the tabulated orbital
elements into Cartesian $(x,y,z)$ positions, which we saved at
12~hour intervals during the calendar years 2018 to 2026.
In order to compare directly with the {\em SPP} trajectory data, each
asteroid's coordinates were interpolated to the denser
(0.1~day) time grid obtained from the spacecraft SPICE kernel.

The specific asteroids appropriate for this study were selected
with the following criteria.
First, we included all 63 known asteroids with perihelia less than
0.2~AU.
The asteroid that approaches closest to the Sun is 2005~HC$_4$,
with perihelion $q = 0.07066$~AU (i.e., 15.2~$R_{\odot}$).
There are six others with perihelia less than 0.1~AU.
Second, the database was extended to asteroids with perihelia between
0.2 and 0.3~AU.
However, out of the 159 known asteroids in this region, many
of them are extremely small and dim.
The results of Sections~\ref{sec:massloss}--\ref{sec:tail} below
show that only the largest asteroids at $q > 0.2$~AU are expected
to display substantial mass loss.
Thus, we selected the brightest 34 of that group, with the criterion
that their $V$-band absolute magnitudes $H$ must be $\leq 18.0$.
No asteroids were chosen with $q > 0.3$~AU because of both their
negligible expected mass loss rates and the infrequency of close
encounters with {\em SPP} once it enters the inner heliosphere.
Table~1 gives the final list of 97 asteroids in order of increasing
perihelion $q$, and it also lists their names/numbers,
eccentricities $e$, and $V$-band magnitudes $H$.

\begin{table}
\renewcommand\thetable{1}
\caption{Inner heliospheric asteroids tracked in this study,
sorted by perihelion.}
\label{tab01}
\begin{tabular}{rlccccc}
\hline\noalign{\smallskip}
\multicolumn{1}{c}{Number} &
\multicolumn{1}{c}{Name} &
$q$ [AU] &
$e$ &
$H$ [mag] &
$d_{\rm min}$ [AU] &
$r_{a} (d_{\rm min})$ [AU] \\
\noalign{\smallskip}\hline\noalign{\smallskip}
 ---    &  2005 HC4    &  0.07066 &  0.96119 & 20.7 &  0.10941 &  0.11916 \\
 ---    &  2008 FF5    &  0.07914 &  0.96539 & 23.1 &  0.28942 &  0.46209 \\
 ---    &  2015 EV     &  0.08003 &  0.96100 & 22.5 &  0.64377 &  0.07659 \\
 394130 &  2006 HY51   &  0.08100 &  0.96884 & 17.2 &  0.32687 &  0.41537 \\
 137924 &  2000 BD19   &  0.09199 &  0.89505 & 17.2 &  0.07756 &  0.10245 \\
 374158 &  2004 UL     &  0.09283 &  0.92670 & 18.8 &  0.16533 &  0.22027 \\
 394392 &  2007 EP88   &  0.09558 &  0.88584 & 18.5 &  0.12213 &  0.21280 \\
 ---    &  2011 KE     &  0.10013 &  0.95502 & 19.8 &  0.15917 &  0.39329 \\
 ---    &  2008 HW1    &  0.10171 &  0.96061 & 17.4 &  0.55013 &  0.25181 \\
 ---    &  2015 HG     &  0.10472 &  0.95025 & 21.0 &  0.35597 &  0.13877 \\
 ---    &  2012 US68   &  0.10566 &  0.95776 & 18.3 &  0.15636 &  0.46061 \\
 ---    &  2011 XA3    &  0.10859 &  0.92597 & 20.5 &  0.26899 &  0.39000 \\
 399457 &  2002 PD43   &  0.11031 &  0.95603 & 19.1 &  0.41708 &  0.48005 \\
 386454 &  2008 XM     &  0.11107 &  0.90913 & 20.0 &  0.13714 &  0.29387 \\
 431760 &  2008 HE     &  0.11337 &  0.94993 & 18.1 &  0.21685 &  0.12394 \\
 ---    &  2007 EB26   &  0.11573 &  0.78867 & 19.6 &  0.01372 &  0.20750 \\
 276033 &  2002 AJ129  &  0.11671 &  0.91488 & 18.7 &  0.36247 &  0.15895 \\
 ---    &  2000 LK     &  0.11788 &  0.94590 & 18.4 &  0.54865 &  0.26888 \\
 425755 &  2011 CP4    &  0.11813 &  0.87039 & 21.2 &  0.12991 &  0.24157 \\
 ---    &  1995 CR     &  0.11931 &  0.86846 & 21.7 &  0.20231 &  0.23316 \\
 ---    &  2007 GT3    &  0.12088 &  0.93938 & 19.7 &  0.55496 &  0.12185 \\
 ---    &  2004 QX2    &  0.12498 &  0.90291 & 21.7 &  0.07061 &  0.69866 \\
 ---    &  2011 BT59   &  0.12859 &  0.94848 & 21.0 &  0.25844 &  0.14597 \\
 289227 &  2004 XY60   &  0.13017 &  0.79669 & 18.9 &  0.07934 &  0.13191 \\
 ---    &  2015 KO120  &  0.13120 &  0.92577 & 22.0 &  0.05460 &  0.67427 \\
 ---    &  2007 PR10   &  0.13241 &  0.89262 & 20.7 &  0.21100 &  0.75243 \\
 ---    &  2006 TC     &  0.13561 &  0.91184 & 18.8 &  0.22618 &  0.23233 \\
 ---    &  2013 JA36   &  0.13750 &  0.94854 & 21.0 &  0.43394 &  0.58412 \\
 ---    &  2008 MG1    &  0.13886 &  0.82271 & 19.9 &  0.03179 &  0.90056 \\
 ---    &  2013 HK11   &  0.13901 &  0.93678 & 20.7 &  0.04258 &  0.15407 \\
   3200 &  Phaethon    &  0.14004 &  0.88984 & 14.6 &  0.24292 &  0.26831 \\
 ---    &  2013 YC     &  0.14104 &  0.94347 & 21.3 &  0.63135 &  0.15442 \\
 ---    &  2010 JG87   &  0.14432 &  0.94773 & 19.1 &  0.48764 &  0.40593 \\
 ---    &  2015 KP157  &  0.14820 &  0.91027 & 19.2 &  0.17604 &  0.57849 \\
 ---    &  2015 DU180  &  0.15228 &  0.92097 & 20.8 &  0.24979 &  0.18051 \\
 ---    &  2012 UA34   &  0.15597 &  0.80155 & 19.5 &  0.05960 &  0.35570 \\
 ---    &  2005 EL70   &  0.15893 &  0.94022 & 24.0 &  0.10759 &  0.36308 \\
 155140 &  2005 UD     &  0.16287 &  0.87224 & 17.3 &  0.06729 &  0.17065 \\
 364136 &  2006 CJ     &  0.16580 &  0.75492 & 20.2 &  0.02326 &  0.21205 \\
 105140 &  2000 NL10   &  0.16727 &  0.81704 & 15.8 &  0.23609 &  0.44102 \\
 ---    &  2011 WN15   &  0.17285 &  0.85793 & 19.6 &  0.17488 &  0.17701 \\
 ---    &  2013 WM     &  0.17466 &  0.91598 & 23.8 &  0.53346 &  0.25236 \\
 302169 &  2001 TD45   &  0.17733 &  0.77742 & 19.9 &  0.12274 &  0.59688 \\
 ---    &  2005 RV24   &  0.17805 &  0.88177 & 20.6 &  0.20295 &  0.75445 \\
 ---    &  2008 EY68   &  0.17888 &  0.75994 & 22.0 &  0.12325 &  0.17417 \\
 141851 &  2002 PM6    &  0.17955 &  0.85012 & 17.7 &  0.08396 &  0.58174 \\
 267223 &  2001 DQ8    &  0.18138 &  0.90150 & 18.0 &  0.06021 &  0.18495 \\
 ---    &  2013 AJ91   &  0.18187 &  0.92818 & 19.3 &  0.18173 &  0.32416 \\
 259221 &  2003 BA21   &  0.18350 &  0.83321 & 19.1 &  0.06899 &  0.23841 \\
 ---    &  2011 YX62   &  0.18409 &  0.92823 & 23.0 &  0.26575 &  0.46486 \\
   1566 &  Icarus      &  0.18652 &  0.82696 & 16.9 &  0.21870 &  0.29748 \\
  89958 &  2002 LY45   &  0.18675 &  0.88625 & 17.0 &  0.14820 &  0.22125 \\
 ---    &  2009 HU58   &  0.18686 &  0.90955 & 19.1 &  0.11882 &  0.34948 \\
   5786 &  Talos       &  0.18727 &  0.82684 & 17.1 &  0.32613 &  0.78989 \\
 ---    &  2003 UW29   &  0.18899 &  0.83840 & 20.7 &  0.06178 &  0.35127 \\
\noalign{\smallskip}\hline
\end{tabular}
\end{table}

\begin{table}
\renewcommand\thetable{1}
\caption{{\bf (continued) \, }
Inner heliospheric asteroids tracked in this study,
sorted by perihelion.}
\begin{tabular}{rlccccc}
\hline\noalign{\smallskip}
\multicolumn{1}{c}{Number} &
\multicolumn{1}{c}{Name} &
$q$ [AU] &
$e$ &
$H$ [mag] &
$d_{\rm min}$ [AU] &
$r_{a} (d_{\rm min})$ [AU] \\
\noalign{\smallskip}\hline\noalign{\smallskip}
 387505 &  1998 KN3    &  0.19537 &  0.87328 & 18.4 &  0.18166 &  0.20196 \\
 ---    &  2007 MK6    &  0.19586 &  0.81879 & 19.9 &  0.29098 &  0.37447 \\
 ---    &  2015 KJ122  &  0.19613 &  0.75026 & 22.0 &  0.06455 &  0.50995 \\
 ---    &  2015 DZ53   &  0.19620 &  0.87013 & 20.8 &  0.40446 &  0.20443 \\
 ---    &  2010 VA12   &  0.19875 &  0.84334 & 19.5 &  0.07376 &  0.22421 \\
 ---    &  1996 BT     &  0.19978 &  0.83500 & 23.0 &  0.10126 &  0.37595 \\
 153201 &  2000 WO107  &  0.19985 &  0.78072 & 19.3 &  0.19124 &  0.35397 \\
 139289 &  2001 KR1    &  0.19996 &  0.84123 & 17.6 &  0.02376 &  0.47075 \\
  66391 &  1999 KW4    &  0.20010 &  0.68846 & 16.5 &  0.12587 &  0.20703 \\
 141079 &  2001 XS30   &  0.20015 &  0.82815 & 17.7 &  0.03979 &  0.20640 \\
 143637 &  2003 LP6    &  0.20341 &  0.88352 & 16.3 &  0.31278 &  0.51759 \\
 329915 &  2005 MB     &  0.20411 &  0.79284 & 17.1 &  0.33105 &  0.54317 \\
 438116 &  2005 NX44   &  0.20495 &  0.90745 & 17.3 &  0.35076 &  0.81328 \\
 369296 &  2009 SU19   &  0.20935 &  0.89942 & 17.9 &  0.06908 &  0.26797 \\
 184990 &  2006 KE89   &  0.21144 &  0.79925 & 16.4 &  0.14092 &  0.52566 \\
 ---    &  2004 LG     &  0.21250 &  0.89714 & 18.0 &  0.39704 &  0.38711 \\
 ---    &  2005 GL9    &  0.22226 &  0.89620 & 17.1 &  0.03307 &  0.22339 \\
 137052 &  Tjelvar     &  0.23768 &  0.80955 & 16.9 &  0.08579 &  0.24021 \\
 225416 &  1999 YC     &  0.24099 &  0.83050 & 17.2 &  0.29616 &  0.31271 \\
 ---    &  2000 SG8    &  0.24508 &  0.90066 & 17.5 &  0.59763 &  0.75514 \\
 242643 &  2005 NZ6    &  0.24872 &  0.86443 & 17.4 &  0.08031 &  0.81175 \\
 136874 &  1998 FH74   &  0.25390 &  0.88462 & 15.7 &  0.29403 &  0.29322 \\
  40267 &  1999 GJ4    &  0.25669 &  0.80825 & 15.4 &  0.31560 &  0.25078 \\
 ---    &  2011 WS2    &  0.25890 &  0.74356 & 17.2 &  0.16082 &  0.32279 \\
 ---    &  2007 VL243  &  0.26200 &  0.72856 & 17.8 &  0.04606 &  0.46376 \\
 331471 &  1984 QY1    &  0.26348 &  0.89453 & 15.4 &  0.28198 &  0.34163 \\
 ---    &  2006 OS9    &  0.26379 &  0.90377 & 17.8 &  0.21360 &  0.35270 \\
 369452 &  2010 LG14   &  0.26898 &  0.74267 & 17.9 &  0.32184 &  0.36409 \\
 190119 &  2004 VA64   &  0.27010 &  0.89042 & 17.1 &  0.41024 &  0.37468 \\
 351370 &  2005 EY     &  0.27510 &  0.89066 & 17.2 &  0.12769 &  0.27657 \\
 164201 &  2004 EC     &  0.28058 &  0.85954 & 15.7 &  0.12414 &  0.28329 \\
 385402 &  2002 WZ2    &  0.28476 &  0.88432 & 17.0 &  0.47899 &  0.34397 \\
 397237 &  2006 KZ112  &  0.28545 &  0.88694 & 16.7 &  0.75831 &  0.66348 \\
 253106 &  2002 UR3    &  0.28549 &  0.79295 & 16.4 &  0.52059 &  0.66113 \\
 364877 &  2008 EM9    &  0.29101 &  0.85153 & 17.3 &  0.08666 &  0.34927 \\
 ---    &  2014 MR26   &  0.29387 &  0.76593 & 17.8 &  0.12435 &  0.66948 \\
 231937 &  2001 FO32   &  0.29523 &  0.82644 & 17.7 &  0.20666 &  0.32020 \\
  99907 &  1989 VA     &  0.29525 &  0.59468 & 17.9 &  0.24233 &  0.31057 \\
 170502 &  2003 WM7    &  0.29648 &  0.88027 & 17.2 &  0.20954 &  0.77615 \\
 ---    &  2010 KY127  &  0.29686 &  0.88116 & 17.0 &  0.51914 &  0.36652 \\
 162269 &  1999 VO6    &  0.29734 &  0.73809 & 17.0 &  0.26966 &  0.34300 \\
 141525 &  2002 FV5    &  0.29916 &  0.72475 & 17.9 &  0.21795 &  0.33586 \\
\noalign{\smallskip}\hline
\end{tabular}
\end{table}

A set of 97 timelines containing the relative positions of {\em SPP,}
the asteroid, and the Sun were produced from the ephemeris data over
the 2018--2025 mission period.
The three mutual distances $d$ (between {\em SPP} and asteroid),
$r_a$ (between Sun and asteroid), and
$r_p$ (between Sun and {\em SPP}) are used for various purposes in
the models described below.
Other useful quantities include the solar elongation angle
$\varepsilon$ (i.e., the angle centered on {\em SPP} between vectors
pointing to the Sun and to the asteroid)
and the scattering phase angle $\alpha$ (i.e., the angle centered on
the asteroid between vectors pointing to the Sun and to
the observer on {\em SPP}).
Both angles can be computed from the three distances,
\begin{equation}
  \cos \varepsilon \, = \,
  \frac{r_{p}^{2} + d^{2} - r_{a}^{2}}{2 r_{p} d}
  \,\,\, , \,\,\,\,\,\,\,\,\,
  \cos \alpha \, = \,
  \frac{r_{a}^{2} + d^{2} - r_{p}^{2}}{2 r_{a} d}
  \,\,\, .
\end{equation}
Table~1 lists $d_{\rm min}$, the distance of closest approach for
each asteroid to {\em SPP}, and the value of $r_a$ at that time.
The asteroid with the smallest value of $d_{\rm min}$ is 2007~EB$_{26}$,
with a minimum separation of only 0.0137~AU (i.e., 2.95~$R_{\odot}$),
which should occur at $t = 779.5$~days.
There are 24 asteroids in the list that come closer than 0.1~AU
to {\em SPP.}

\begin{figure}
\hspace*{0.20in}
\includegraphics[width=0.93\textwidth]{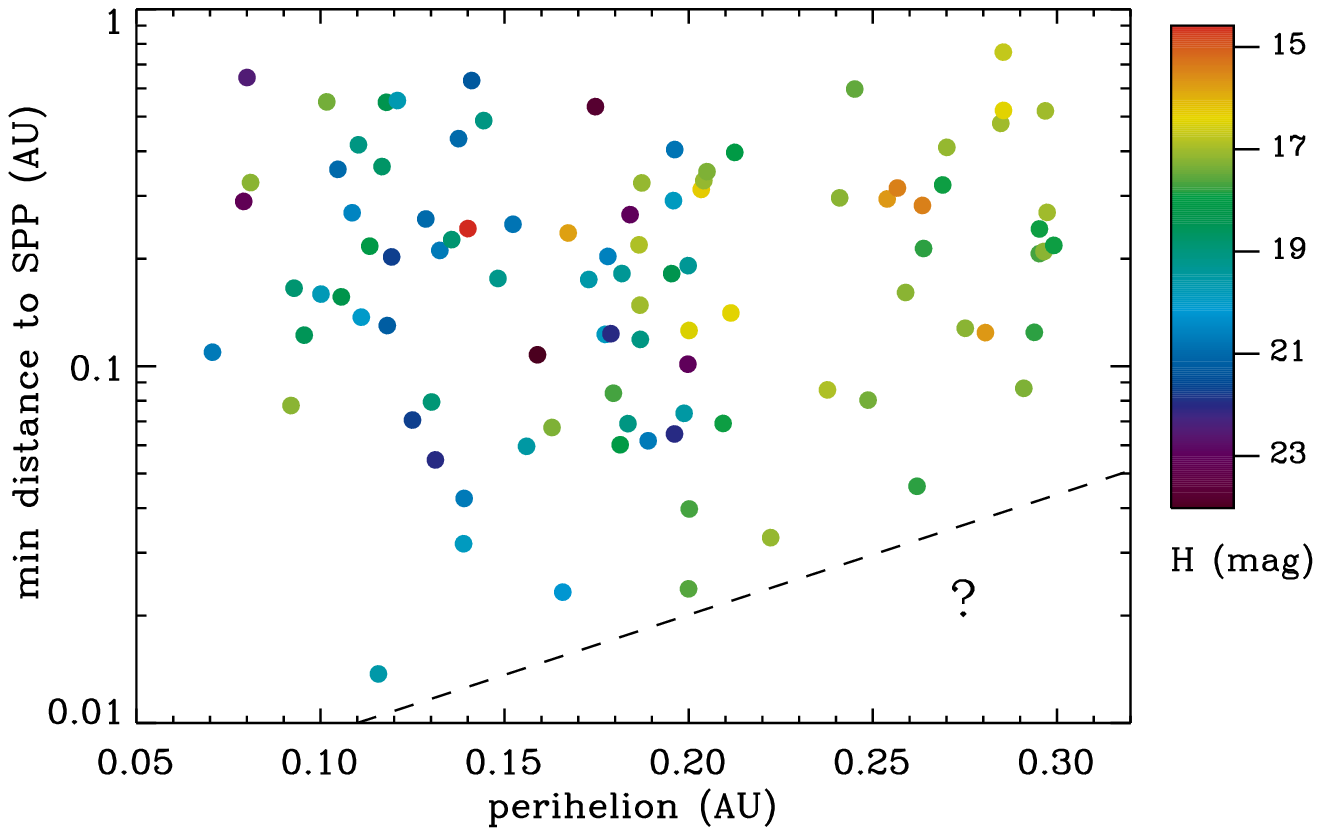}
\caption{Minimum distances $d_{\rm min}$ between {\em SPP} and each
asteroid, found by comparing ephemerides between July 31, 2018 and
September 1, 2025.  The symbol colors map to asteroid absolute
magnitude $H$.}
\label{fig02}
\end{figure}

Figure~\ref{fig02} shows the spread of minimum distance
$d_{\rm min}$ versus each asteroid's perihelion distance $q$.
There is no overall correlation between these quantities, but there does
appear to be a slight preponderance (see dashed line) for asteroids with
the largest $q$ values to avoid close approaches with {\em SPP.}
This is likely to be a statistical trend associated with the fact
that asteroids with wider orbits naturally spend less time in
{\em SPP's} neighborhood close to the Sun.

It should be emphasized that the accurate prediction of a close approach
between {\em SPP} and any specific asteroid depends on the validity
of the planned trajectory and launch date of July 31, 2018.
Spacecraft launches are frequently delayed, but {\em SPP} does have
firm requirements to ``meet'' the desired gravitational assists
with Venus.
Thus, it is possible that even a delayed launch could result in an
eventual synchronization with the trajectory assumed here.
In any case, some results of this paper may be better interpreted
as one possible sample from a quasi-random distribution of possible
{\em SPP} orbits.
Additional statistical conclusions are discussed in
Section~\ref{sec:conc}.

\begin{figure}
\hspace*{0.20in}
\includegraphics[width=0.85\textwidth]{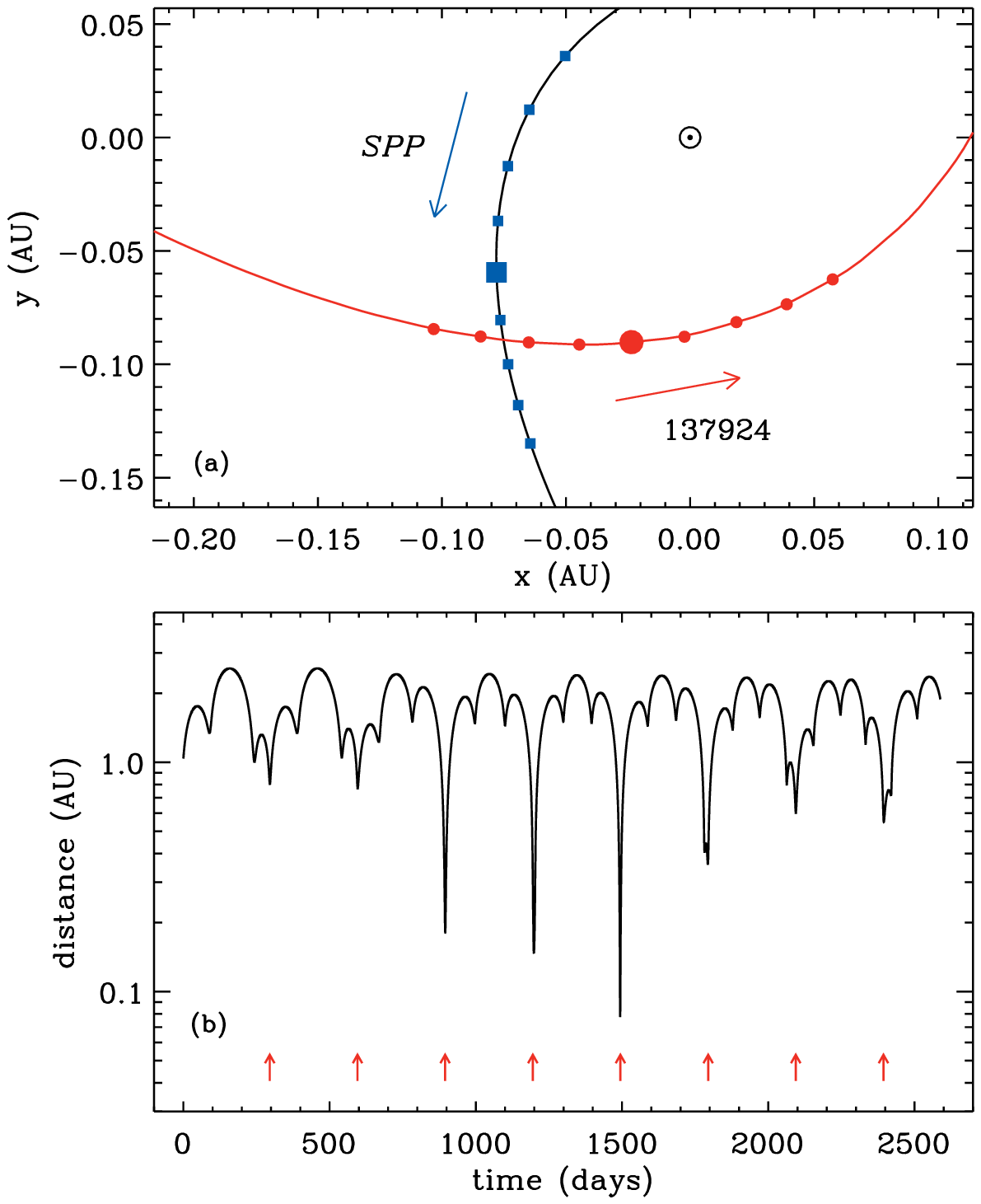}
\caption{(a) Trajectories for the closest encounter between {\em SPP}
(black curve) and asteroid 137924 (red curve).
Large symbols show positions at minimum distance, at mission time
1493.68 days.
Smaller symbols show the positions along a sequence of 7.2~hr intervals
before and after the time of minimum distance.
(b) Relative distance between {\em SPP} and asteroid 137924
versus mission time in days.
Red arrows show the times of the asteroid's perihelia.}
\label{fig03}
\end{figure}

As an example of an interesting encounter, Figure~\ref{fig03}(a)
shows the mutual trajectories for {\em SPP} and asteroid 137924
(2000~BD$_{19}$).
This example is only the 19th closest approach out of the full list
of 97 asteroids, but it is notable for occurring very near the
asteroid's own perihelion.
That fact leads to it having a bright $V$ magnitude and unusually
high predicted values for its mass loss rate and coma size
(see below).
Figure~\ref{fig03}(b) also shows that, for the case of asteroid
137924, multiple close encounters with {\em SPP} tend to occur
near successive perihelion passes.

\section{Asteroid Physical Properties}
\label{sec:props}

In the analysis performed below, each asteroid is assumed
to be spherical in shape, with a diameter $D$ given by a standard
conversion from its visible-light absolute magnitude $H$,
\begin{equation}
  D \, = \, \left( 1348 \, p^{-1/2} \, 10^{-H/5} \right)
  \,\,\,\,\, \mbox{km} \,\, ,
\end{equation}
\citep[e.g.,][]{Bw89},
where we take $p=0.1$ as a representative value of the geometric
albedo for small asteroids in the inner heliosphere
\citep[see also][]{Mu95,Mo02}.
For the 97 asteroids listed in Table~1, the median derived value
of $D$ is 0.890 km, and the minimum and maximum values are
0.0675 km (2005 EL$_{70}$) and 5.12 km (3200 Phaethon), respectively.
Their angular sizes, as seen by {\em SPP}, are typically in the range
between 0.001$''$ and 0.03$''$, with the largest value of 0.075$''$
found for the closest approach of asteroid 2001 KR$_1$.
These values are much smaller than the spatial resolution of the
WISPR telescopes.
However, we anticipate that many asteroids will emit bright dust
clouds that extend to distances several orders of magnitude larger
than their respective diameters (see Section \ref{sec:tail}).

\begin{figure}
\hspace*{0.20in}
\includegraphics[width=0.85\textwidth]{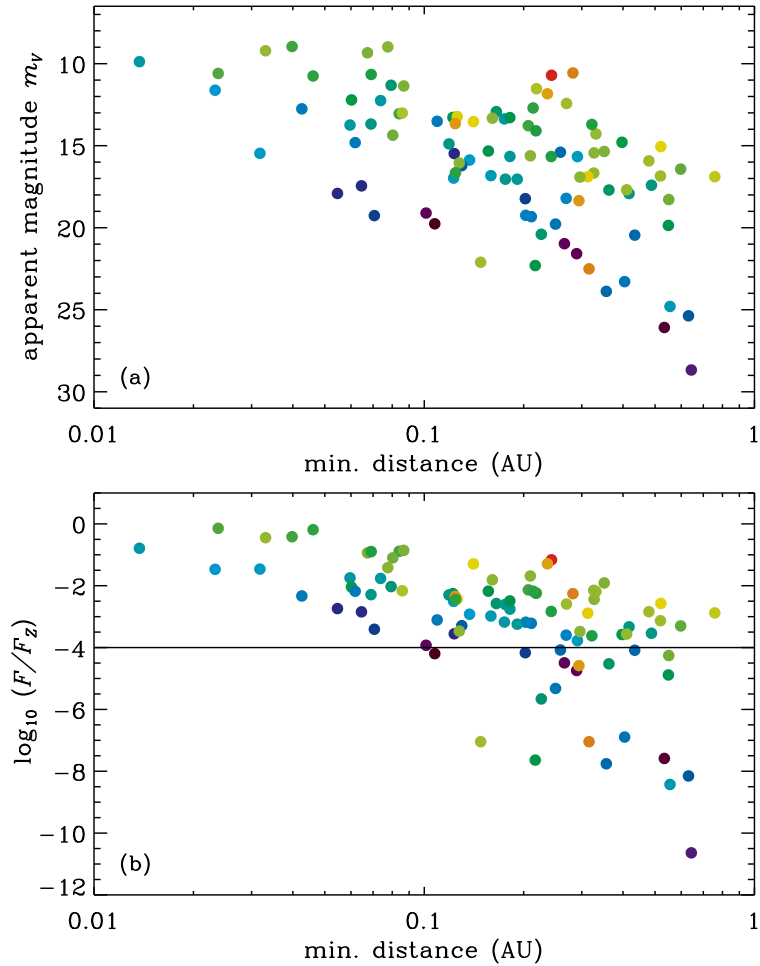}
\caption{(a) Apparent magnitudes, at times of closest approach with
{\em SPP,} plotted versus minimum distance between each asteroid
and {\em SPP.}
(b) $V$-band asteroid flux $F$ divided by the estimated zodiacal
light flux $F_Z$ measured in a single WISPR pixel.
The horizontal line shows an approximate observable threshold of
$10^{-4}$ times the background.
In both panels, symbol colors correspond to $H$ with the same scaling
as in Figure~\ref{fig02}.}
\label{fig04}
\end{figure}

The apparent magnitude of an asteroid in the Johnson--Cousins $V$
band (dominated by wavelengths between 500 and 600 nm) is given by
\begin{equation}
  m_{V} \, = \, H + 5 \log_{10} ( r_{a} d ) - 2.5 \log_{10} \Phi(\alpha)
\end{equation}
where $r_a$ and $d$ are given in units of AU.
$\Phi(\alpha)$ is the phase function for the scattering of sunlight,
which is largest at opposition ($\alpha = 0$) and decreases
monotonically for larger scattering angles.
The standard phase function defined by \citet{Bw89} was used with a
slope parameter $G=0.15$ appropriate for inner heliospheric
asteroids \citep[see also][]{LM90}.
Apparent magnitudes were converted into $V$-band energy fluxes
at $\lambda \approx 500$~nm, with
\begin{equation}
  F \, = \, 3.67 \times 10^{-23} \, ( 10^{-0.4 m_{V}} )
  \,\,\,\, \mbox{W m$^{-2}$ Hz$^{-1}$}
\end{equation}
using the normalization factor specified by \citet{Wa81}.
Figure~\ref{fig04}(a) shows the distribution of $m_V$ versus
minimum distance $d_{\rm min}$.
The brightest one is asteroid 141079, with $m_{V} = 8.95$,
but there is nothing unusual about its intrinsic properties.
Finding a small value of $m_V$ depends on the chance of finding small
values for $d$, $r_a$, and $\alpha$, all occurring at roughly
the same time.

The observability of a given asteroid depends on both its intrinsic
brightness and its relative contrast with the sky background.
\citet{Je13} observed 3200 Phaethon and its tail with the
Sun Earth Connection Coronal and Heliospheric Investigation (SECCHI)
package on {\em STEREO} \citep{Ho08,Ey09},
and they noted how measurements were limited by the presence
of extended sky emission.
The WISPR instrument on {\em SPP} will be closer to its asteroid
targets than was SECCHI, and it will have comparable sensitivity
despite its smaller size.
Thus, in this paper we apply several of the lessons learned from
SECCHI to future measurements with WISPR.

Visible-light sky emission in the inner heliosphere is dominated by
a combination of zodiacal light (the dust-scattered F corona) and
Thomson-scattered electron emission (the K corona).
A relatively simple analytic expression for the total specific
intensity $B_{Z} (\varepsilon, r_{p})$ of the two components was
found to reproduce a range of observations and model predictions.
For a standard observer at $r_{p} = 1$~AU, the angular dependence is
given by
\begin{equation}
  \frac{B_{Z} (\varepsilon, \mbox{1 AU})}{B_{\odot}} \, = \,
  \frac{4 \times 10^{-25}}{\varepsilon^8} +
  \frac{6.3 \times 10^{-14}}{[\sin (0.63\varepsilon)]^{2.22}}
  \label{eq:zodi}
\end{equation}
where the elongation angle $\varepsilon$ is specified in radians and
the mean solar-disk brightness is
$B_{\odot} = 2.96 \times 10^{-8}$ W m$^{-2}$ sr$^{-1}$ Hz$^{-1}$
in the $V$ band \citep{Allen}.
Equation (\ref{eq:zodi}) matches collected observations in the
ecliptic plane \citep{Ln75,MJ77,Kw04,Mn04} to within about a
factor of two.
Observers closer to the Sun are expected to see higher intensities
\citep[e.g.,][]{vD88}, and we use
\begin{equation}
  B_{Z} (\varepsilon, r_{p}) \, = \,
  B_{Z} (\varepsilon, \mbox{1 AU}) \, \left(
  \frac{\mbox{1 AU}}{r_p} \right)^{2.5}
\end{equation}
where the exponent 2.5 was estimated from the modeled inner
heliospheric intensities in Figure~7 of \citet{WISPR}.

The goal is to compare the above sky brightness with the flux from
an asteroid, but the latter is essentially a tiny point-source with an
angular size much smaller than a WISPR detector pixel.
The asteroid's flux $F$ must be compared with a corresponding
sky flux $F_Z$ that fills the pixel.
The zodiacal light intensity $B_Z$ is thus converted into flux by
multiplying it by the solid angle subtended by a pixel, with
\begin{equation}
  F_{Z} \, = \, \Omega \, B_{Z}(\varepsilon, r_{p}) \,\, .
\end{equation}
\citet{WISPR} specified pixel sizes of 1.2$'$ and 1.7$'$ for the inner
and outer fields of view, and we used the average of the two as a
representative value.
Thus, a square patch of the sky subtending
$1.45' = 4.22 \times 10^{-4}$ radians on a side occupies
a solid angle $\Omega = 1.78 \times 10^{-7}$ sr.

Figure~\ref{fig04}(b) shows the ratio $F/F_{Z}$ as a function
of $d_{\rm min}$.
Differences with panel (a) are mainly due to the fact that different
asteroids are viewed with different elongation angles, so the sky
brightness varies.
The highest-contrast asteroid is 2001~KR$_1$, with $F/F_{Z} = 0.72$.
Ideally, one would consider a ``good'' observation to be one with
$F \gg F_{Z}$, but experience with SECCHI has shown that much weaker
signals can be extracted from strong backgrounds.
\citet{Df11} used sophisticated processing techniques to resolve
features with fluxes as low as $(F/F_{Z}) \approx 10^{-4}$.
That level is indicated with a horizontal line in Figure~\ref{fig04}(b),
and we note that 76 out of 97 asteroids fall above that level.
WISPR will clearly have multiple opportunities to observe
asteroids in the inner heliosphere.

It is noteworthy that the list of five brightest (i.e., lowest $m_V$)
asteroids and the list of five highest flux contrast (largest $F/F_Z$)
asteroids share three members: 141079, 2005~GL$_{9}$, and 2007~EB$_{26}$.
Note that asteroid 137924, whose orbit is shown in
Figure~\ref{fig03}, is the second brightest ($m_{V} = 8.98$) at
closest approach, but due to a low elongation angle it is only the
14th highest flux contrast ($F/F_{Z} = 0.038$).
The largest active asteroid, 3200 Phaethon, has its lowest value
of $m_V = 10.53$ at $t = 1384.6$, about six days prior to the time it
reaches its minimum distance $d_{\rm min}$ to {\em SPP.}
However, because of varying elongation angles, Phaethon's time of
maximum flux contrast (at $F/F_{Z} = 0.13$) occurs at its previous
perihelion passage 17 months earlier ($t = 880.3$).
This tells us that the ``snapshots'' of the 97 closest-approach events,
illustrated by filled circles in Figures \ref{fig02} and \ref{fig04},
may not always be the most reliable guide to identifying when
interesting things are occurring.

\section{Asteroid Mass Loss due to Erosion}
\label{sec:massloss}

In a similar manner to comets, it is believed that active asteroids
lose mass when they orbit sufficiently close to the Sun.
The ejected gas and dust expands to fill coma-like or tail-like
atmospheres that may be observed at large distances from the nucleus.
One can use the language of sublimation to discuss the erosion of
solid material from the asteroid surface.
Some active asteroids \citep[e.g., 133P/Elst-Pizzaro in the main asteroid
belt; see][]{Hs04} are probably similar to comets in that they emit
gaseous mass mostly in the form of volatile ices (H$_2$O, CO, CO$_2$).
However, the near-Sun active asteroids (e.g., Phaethon) have likely
already lost most of these easily sublimated compounds.
The heavier silicate and hydrocarbon molecules that presumably dominate
the outer regolith layers of these asteroids have a substantially
higher latent heat than volatile comet ices.
This section applies the long history of sublimation energy-balance
models for comets \citep[e.g,][]{Wg59,DM71,WH76,We80,Pr04}
to the case of near-Sun active asteroids.

Before proceeding with such modeling, it is important to note that
the loss of both silicate-rich gas and dust from rocky objects (heated to
$T \approx 1000$~K) has not been studied as extensively as the standard
cometary scenario of sublimating ice molecules that drag along the
larger dust grains.
However, the idea of a simultaneous ejection of multiple phases of the
{\em same} type of material has been studied for comets.
Water ice and other volatile compounds appear to be ejected in both the
gas phase and in the form of ``snow'' particles with sizes ranging from
microns to meters \citep[e.g.,][]{DM71,Ke13,Ke15,Pp14}.

In addition to the analogy with comet ice loss, there are three
other comparable situations that appear to have some resemblance and
relevance to the case of active asteroids:
\begin{enumerate}
\item
Meteors entering a planetary atmosphere undergo rapid deceleration
and thermal ablation \citep[e.g.,][]{BS74}.
Although the source of heating for meteors is different
(atmospheric drag versus solar radiation), the mass loss is
often treated by replacing the latent heat of sublimation with a
comparable heat of ablation \citep{Ch93}.
The size distribution of resulting fragments appears to be quite broad,
from nanometer-scale ``smoke'' to micron-scale dust to macroscopic
meteoroids \citep{BC09,MM11}.
\item
Sungrazing comets have perihelia within a few solar radii of the Sun's
surface \citep{Ma05}, and observations of their evolution upon close
approach provide valuable information about the composition and
thermal properties of primordial bodies in the solar system.
\citet{Sk03} was able to model the light curves of a number of sungrazing
comets by assuming a continuous distribution of silicate erosion products
(i.e., from large fragments to individual sublimated molecules).
These comets have also been observed to emit metallic---i.e.,
alkali, sulfide-rich, and iron-group---material as well
\citep{Pr67,Sl69,Zo06,Ci10},
\item
We now know that a large number of extrasolar planets orbit within
just a few stellar radii of their host stars \citep{WF15}.
Some observations suggest that small, rocky planets of this kind are
slowly disintegrating via the ejection of large amounts of dust
\citep[e.g.,][]{Mu11,Rp12,vL14,SO15}.
Possible formation channels for the dust include condensation from
sublimated gas, direct ejection via volcanism, or comet-like
entrainment of grains along with escaping gas molecules.
\end{enumerate}
Thus, there appear to be multiple ways that silicate-rich gas can be
removed from the surfaces of solid objects near the Sun, together with
larger dust grains, when they experience strong solar irradiation.
It should also be mentioned that the minerals left on the surfaces of
these bodies may be irradiated sufficiently to induce various types
of chemical and tensile metamorphosis \citep{Sc89,Sch5,Ks06,GB16}.

The thermal energy balance at the surface of a solid body
can be solved to compute the mass loss rate $\dot{M}$ and equilibrium
temperature $T$ of molecules leaving the surface.
Ignoring heat conduction, the short-wave radiative energy gained
must be balanced by losses due to long-wave radiation and sublimation,
with
\begin{equation}
  \frac{(1-A) S_{\odot}}{r_{a}^2} \cos\theta
  \, = \, \eta \sigma_{\rm B} T^{4} \, + \, m_{u} L Z
  \label{eq:sek}
\end{equation}
where $A$ is the asteroid's Bond albedo, $S_{\odot}$ is the solar
constant, $\theta$ is the angle between rays from the Sun and the
asteroid surface normal, $\sigma_{\rm B}$ is the Stefan-Boltzmann
constant, $\eta$ is the emissivity of the asteroid surface,
$m_u$ is the atomic mass unit,
$L$ is the latent heat of sublimation of the escaping gas,
specified here as an energy per mole,
and $Z$ is the surface sublimation rate
(particles lost per unit area per unit time).
For simplicity, some of these quantities are fixed at constant values
of $A=0.1$, $\eta = 1$, and $S_{\odot} = 1360$~W m$^{-2}$.

Although the resulting rate of mass loss is sensitive to details of
the Sun--asteroid geometry, it is possible to discuss the behavior
in two limiting cases for $\cos\theta$ \citep[see][]{Je11,Je15}.
A {\em rapidly rotating} asteroid will redistribute the incoming
radiative energy from its Sun-facing side over the majority of its
emitting surface.
In that case, a mean value of $\cos\theta = 1/4$ accounts for this
efficient redistribution.
On the other hand, a {\em slowly rotating} (or non-rotating)
asteroid will receive most of its energy at nearly sub-solar
surface locations, and will emit sublimated gas only from the
illuminated parts.
Thus, one can assume $\cos\theta = 1$ at those locations.
In these two cases the emitting surface area $A$ of the asteroid
also differs.
In the fast-rotating limit, nearly all points on the asteroid
receive some heat, so $A \approx \pi D^2$ (the full surface area).
In the slow-rotating limit, the area is given roughly by the
front-side cross section, $A \approx \pi D^2/4$.
This area is needed to compute the full mass loss rate
\begin{equation}
  \dot{M} \, = \, Z \mu m_{u} A \,\, ,
  \label{eq:Mdot}
\end{equation}
where $\mu$ is the mean molar mass of the sublimating molecules.
The quantity $Z \mu m_{u}$ is sometimes called the mass
erosivity $\dot{\epsilon}$ (mass lost per unit time per unit area).

Equation (\ref{eq:sek}) has two undetermined parameters: $T$ and $Z$.
These can be found by using the Clausius--Clapeyron relation,
which specifies the vapor pressure $P_v$ at the surface of a
sublimating body,
\begin{equation}
  P_{v} \, = \, P_{\infty} \, \exp \left(
  - \frac{m_{u} L}{k_{\rm B} T} \right)
  \label{eq:clausius}
\end{equation}
where $P_{\infty}$ is the vapor pressure in the high-temperature
limit and $k_{\rm B}$ is Boltzmann's constant.
The latent heat is assumed to be a constant, independent of temperature.
Assuming a Maxwellian distribution of escaping molecules, the 
mean sublimation rate at the surface is given by the Hertz--Knudsen
equation,
\begin{equation}
  Z \, = \, \frac{\gamma \, P_v}{\sqrt{2\pi \mu m_{u} k_{\rm B}T}}
  \label{eq:zmax}
\end{equation}
where $\gamma$ is a dimensionless efficiency that takes account of
a range of kinetic effects \citep[see, e.g.,][]{vL14}.

\begin{table}
\renewcommand\thetable{2}
\caption{Sublimation properties of cometary and asteroidal substances.}
\label{tab03}
\begin{tabular}{lrrrrcc}
\hline\noalign{\smallskip}
Substance & $\mu$ & $\rho$ &
$L$ & $\ln P_{\infty}$ & $\gamma$ & Refs \\
  & [g/mol] & [g/cm$^3$] &
[kJ/mol] & [ln Pa] &  &  \\
\noalign{\smallskip}\hline\noalign{\smallskip}

Carbon dioxide (CO$_2$) &
  44.00 & 1.56 &  26.34 & 27.84 & 1.0 & 1 \\
Water ice (H$_2$O) &
  18.00 & 1.00 &  48.06 & 28.90 & 1.0 & 2 \\
Sodium (Na) &
  23.00 & 0.97 & 105.80 & 22.53 & 1.0 & 3 \\
Carbonaceous chondrite &
  39.85 & 2.80 & 320.67 & 24.48 & 1.0 & 4 \\
Iron (Fe) &
  55.85 & 7.87 & 402.03 & 26.90 & 1.0 & 5 \\
Silicon monoxide (SiO) &
  44.09 & 2.13 & 411.72 & 30.20 & 0.04 & 5 \\
Fayalite (Fe$_2$SiO$_4$) &
 203.77 & 4.39 & 501.99 & 35.40 & 0.1 & 5 \\
Enstatite (MgSiO$_3$) &
 100.39 & 3.20 & 572.92 & 35.80 & 0.1 & 5 \\
Forsterite (Mg$_2$SiO$_4$) &
 140.69 & 3.27 & 542.99 & 31.80 & 0.1 & 5 \\
Quartz (SiO$_2$) &
  60.08 & 2.60 & 577.38 & 30.80 & 1.0 & 5 \\
Corundum (Al$_2$O$_3$) &
 101.96 & 4.00 & 643.24 & 37.00 & 0.1 & 5 \\
Silicon carbide (SiC) &
  40.10 & 3.22 & 652.36 & 35.50 & 0.1 & 5 \\
Graphite (C) &
  12.01 & 2.16 & 778.60 & 34.40 & 0.1 & 5 \\
\noalign{\smallskip}\hline
\end{tabular}\\
{\footnotesize {\bf References:}
1.\  \citet{FS87},
2.\  \citet{FS84},
3.\  \citet{Hu70},
4.\  \citet{BS74},
5.\  \citet{vL14}.}
\end{table}

\begin{figure}
\hspace*{0.25in}
\includegraphics[width=0.85\textwidth]{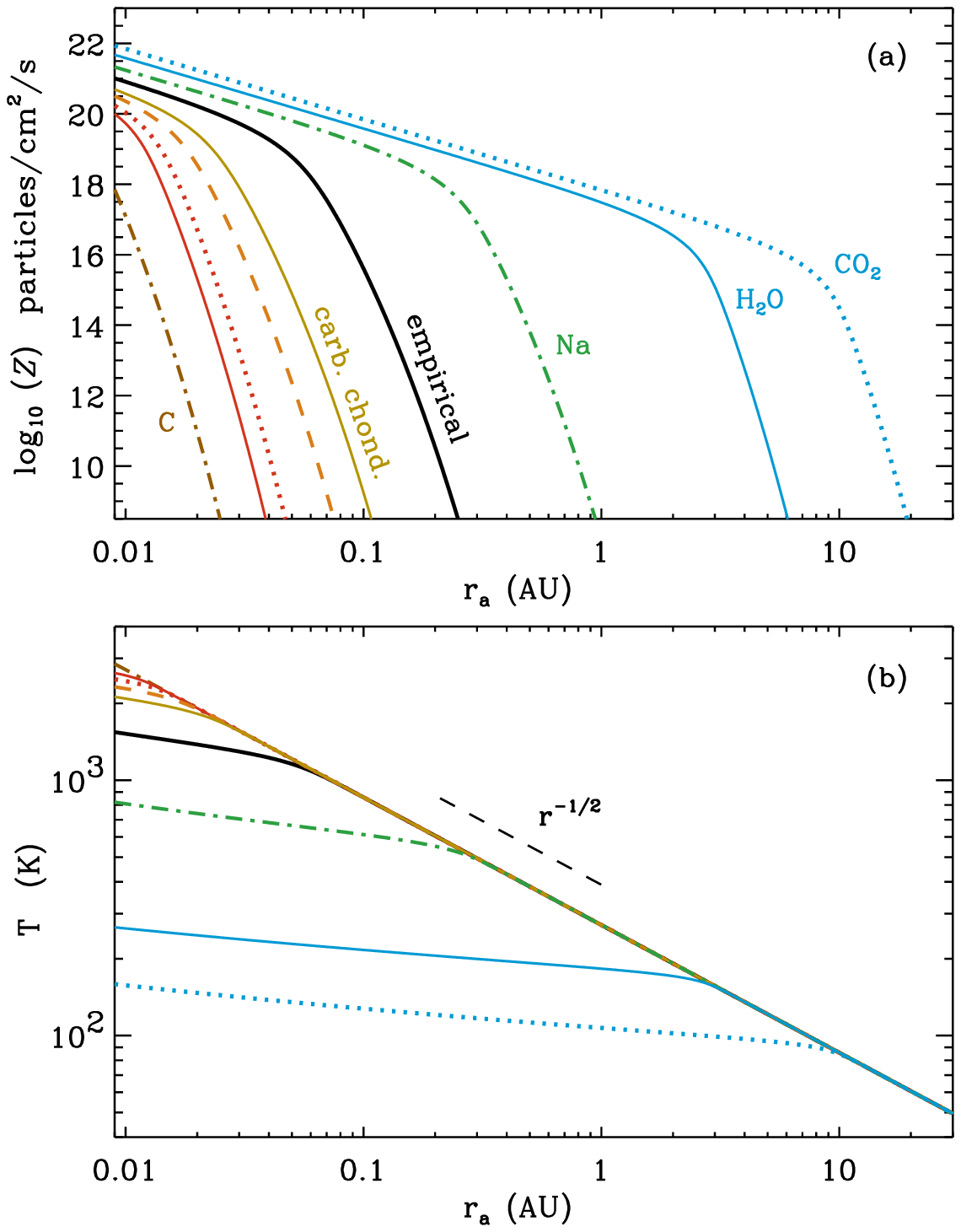}
\caption{Dependence on asteroid heliocentric distance $r_a$ of:
(a) surface sublimation rates $Z$ and
(b) energy-balance temperatures $T$,
plotted for a subset of the substances listed in Table~2.
Unlabeled curves correspond to atomic iron (orange dashed curve),
crystalline enstatite (red dotted curve), and
corundum (red solid curve).  See text for discussion of the
empirical model (black solid curve).}
\label{fig05}
\end{figure}

Table~2 lists many of the above properties for substances believed
to exist on comets and asteroids, listed in order of increasing
latent heat.
Figure~\ref{fig05}(a) shows numerical solutions for $Z$ as a
function of heliocentric distance $r_a$, for a subset of the
substances listed in Table~2.
In most cases there is an inverse monotonic relationship between
$L$ and $Z$ at any given distance.
Close to the Sun, the sublimation term dominates the right-hand side
of Equation (\ref{eq:sek}), and thus $Z \propto r_{a}^{-2}$.
At larger distances, $Z$ drops off exponentially when the
radiative emission term begins to dominate.
The fast-rotating limit was used to compute $\cos\theta$ and $A$
because observations \citep[e.g.,][]{Cp09} suggest inner heliospheric
asteroids experience significantly more surface redistribution of
thermal energy than predicted by non-rotating models.

Figure~\ref{fig05}(b) shows the associated energy-balance solutions
for the surface temperature $T$.
For the most volatile substances, $T$ is relatively flat near the Sun
because the incoming solar flux drives the sublimation phase change
and does not heat up the asteroid.
In that case, the quantity $1/T$ varies approximately as
$(c_{1} + c_{2} \ln r_{a})$ for appropriate constants $c_1$ and $c_2$.
Farther from the Sun, however,
$T \propto r^{-1/2}$ in radiative equilibrium.

\begin{figure}
\hspace*{0.20in}
\includegraphics[width=0.89\textwidth]{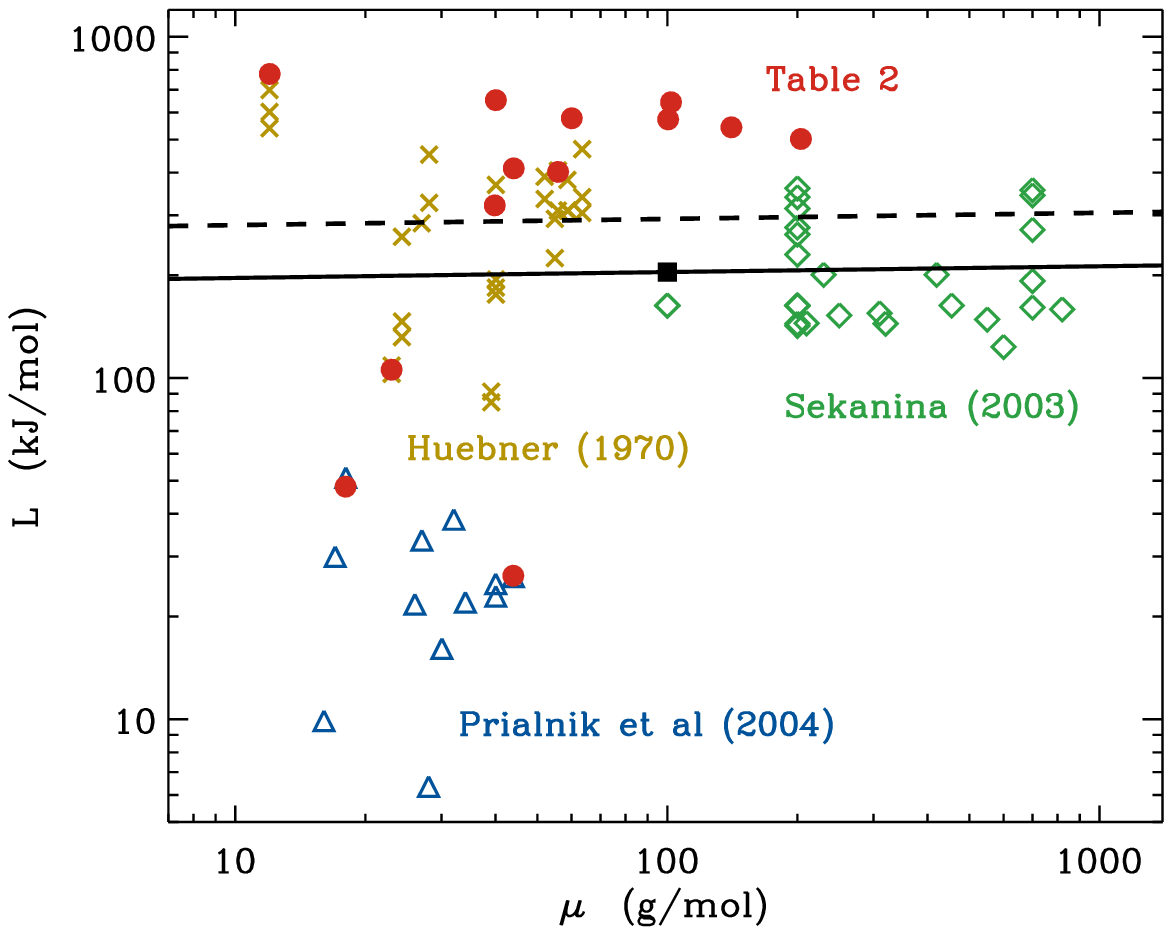}
\caption{Comparison of laboratory measurements of the sublimation
properties of solid substances (red filled circles from Table~2;
gold crosses from \citeauthor{Hu70} \citeyear{Hu70};
blue triangles from \citeauthor{Pr04} \citeyear{Pr04})
with empirical determinations from sungrazing comet light curves
(green diamonds from Tables 1 and 3 of
\citeauthor{Sk03} \citeyear{Sk03}).
Contours that reproduce the measured mass loss of Phaethon are
shown in black (solid: fast-rotating limit, dashed: slow-rotating limit).}
\label{fig06}
\end{figure}

Because we do not yet know the detailed surface composition of active
asteroids, it is not clear how to specify the latent heat and
other properties {\em a~priori.}
\citet{Sk03} modeled the light curves of sungrazing comets by treating
the latent heat $L$ and mean molecular mass $\mu$ as constrained
free parameters.
The result of that process was a range of latent heats (120 to
360 kJ~mol$^{-1}$) and molecular masses (200 to 800 g~mol$^{-1}$),
with unique pairs of values that successfully predict the light
curve properties of each comet.
Figure~\ref{fig06} compares laboratory values of $L$ and $\mu$ for
the substances listed in Table~2 with those derived empirically by
\citet{Sk03}.
Shown for comparison are data from other listings of volatile ices
\citep{Pr04} and solid-phase elements \citep{Hu70}.
In this two-dimensional plane, there is no real overlap between the
laboratory values and those derived by \citet{Sk03}.
However, it is important to note that the sublimation rate $Z$ is
driven mainly by the value of $L$ and only very weakly by $\mu$
(i.e., $Z \propto \mu^{-1/2}$).
Thus, the overlap that is seen in just the $L$ values may point to
the existence of a {\em heterogeneous mixture} of substances with
a range of sizes and latent heats (e.g., hydrated silicates mixed
with organic hydrocarbons) that could be released from surfaces
exposed to the near-Sun environment.

In a similar vein as the empirical study of \citet{Sk03}, it is
possible to use the \citet{Je13} measurement of Phaethon's mass
loss ($\dot{M} \approx 3$ kg~s$^{-1}$ at perihelion) to provide
an observation-based estimate of $L$ and $\mu$.
However, the measured value of $\dot{M}$ from Phaethon was a
dust mass loss rate, whereas the quantity computed in
Equation~(\ref{eq:Mdot}) corresponds to the gas/molecular component
of the outflow.
As stated above, there is still no firmly accepted understanding of
how active asteroids produce and eject dust grains.
The gas and dust components may be intimately connected to one
another (via, e.g., re-condensation of sublimated molecules) or they
may originate from completely different regions on the surface
\citep[see][]{Je15}.
However, we can note that many comets are inferred to have
dust-to-gas mass ratios ${\cal M}$ centered around unity
\citep[e.g.,][]{AH95,Sa96,Ko07}.
Thus, we make a trial assumption of ${\cal M}=1$, which allows us
to take $\dot{M}$ to be equal to the dust mass loss rate.
The assumed value of ${\cal M}$ is likely to be an important source
of uncertainty for the model results given below.
The sensitivity of these results to changes in ${\cal M}$ is
explored further below as well.

An empirical model of the sublimation of material with arbitrary
$L$ and $\mu$ requires specification of the other sublimation properties
of the material.
We follow \citet{Hu70} and \citet{Sk03} by defining
\begin{equation}
  P_{\infty} \, = \, P_{0} \, \exp \left(
  \frac{m_{u} L}{k_{\rm B} T_0} \right)
\end{equation}
with a fiducial value of $P_{0} = 0.398$~GPa.
The result of computing $T_0$ for the 13 actual substances listed
in Table~2 is a mean value (after excluding CO$_2$ and H$_2$O)
of 5300~K, which is used in the models below.\footnote{%
This value is also consistent with the Clausius--Clapeyron relation
used by \citet{Sk03}.}
Since the assumed heterogeneous mixture of substances may contain
both dusty ices ($\gamma \approx 1$) and silicates ($\gamma \approx 0.1$)
we assume a mean value of $\gamma = 0.3$.

A comprehensive search of the two-dimensional space shown in
Figure~\ref{fig06} was undertaken for a Phaethon-like asteroid at
perihelion, with $r_{a} = 0.14004$~AU and $D = 5.12$~km.
Values of $L$ and $\mu$ that yielded $\dot{M} = 3$ kg~s$^{-1}$
are shown by the black curves in Figure~\ref{fig06}; one computed for
the fast-rotating limit and one for the slow-rotating limit.
For simplicity, we choose one point along the fast-rotating locus of
solutions for use in the models presented below:
$L = 204$ kJ~mol$^{-1}$ and $\mu = 100$ g~mol$^{-1}$
(see the black filled square in Figure~\ref{fig06}).
This value of $L$ falls comfortably within the empirically determined
range found by \citet{Sk03} for sungrazing comets.
Also, our solutions are close to the value of 320 kJ~mol$^{-1}$
listed in Table~2 for carbonaceous chondrites.
It should be noted that the value of $L$ for chondritic
material is not really known so precisely.
Both \citet{BS74} and \citet{Ch93} reported approximate values for
the heat of ablation of chondrite meteors spanning values between
200 and 350 kJ~mol$^{-1}$.
To repeat, we believe this empirical solution may point to the
existence of a mixture of multiple solid species that, taken in bulk,
sublimate at a similar rate as a single compound with representative
values of $L$ and $\mu$.

\begin{figure}
\hspace*{0.20in}
\includegraphics[width=0.89\textwidth]{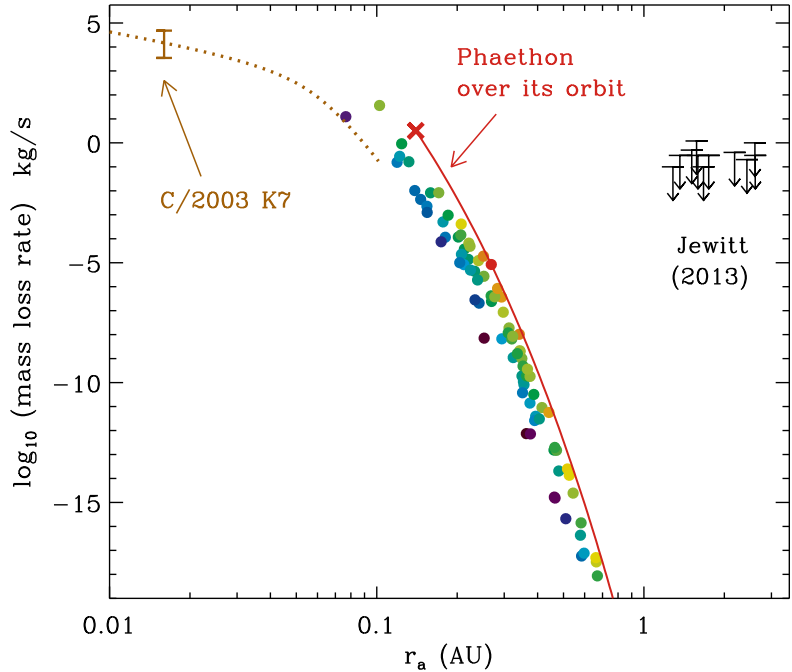}
\caption{Gas mass loss rates $\dot{M}$ for asteroids at their
times of closest approach with {\em SPP} (symbols, with colors
corresponding to $H$ with same scaling as in Figure~\ref{fig02})
and for asteroid 3200 Phaethon at various points along its orbit
(red solid curve) and at perihelion (red X symbol).
A similar model for comet C/2003~K7 (brown dotted curve) is compared
to the measured range of dust $\dot{M}$ values from \citet{Ci10}
(brown strut).
Non-detection upper limits from Keck observations of active asteroids
far from perihelion \citep{Jkeck} are shown with black arrows.}
\label{fig07}
\end{figure}

Figure~\ref{fig07} shows computed mass loss rates versus $r_a$ for
this paper's collection of 97 closest-approach events with {\em SPP,}
and we use the empirical solution described above:
$L = 204$ kJ~mol$^{-1}$ and $\mu = 100$ g~mol$^{-1}$ in the
fast-rotating limit.
When these parameters are held fixed, it is clear that $r_a$ is the
parameter most important for determining $\dot{M}$,
with the asteroid diameter $D$ providing a much more limited variation.
The red filled circle indicates the position of asteroid Phaethon at
its closest approach with {\em SPP,} and the red X symbol shows
Phaethon at its perihelion.
At the time of closest approach with {\em SPP,} Phaethon's mass loss
rate is predicted to be almost a million times lower than when it
reaches perihelion.
This again provides a warning that the times of closest approach
may not be the most auspicious observation times.
Observations of Phaethon and several other asteroids taken at larger
distances \citep[$r_{a} = 1.4$--2.6~AU; see][]{Jkeck} showed no
discernible mass loss.
The approximate upper limits on $\dot{M}$ shown in Figure~\ref{fig07}
are many orders of magnitude larger than what our model would predict
at those distances.

Another comparison between observed dust mass loss and the model
described above is shown in Figure~\ref{fig07}.
\citet{Ci10} observed silicon and carbon ultraviolet emission from
sungrazing comet C/2003~K7 when it was at a distance of
$r_{a} = 0.016$~AU.
They deduced a range of dust mass loss rates assuming the silicon
comes from the dissociation of molecules like olivine or forsterite.
\citet{Ci10} also inferred a diameter for the nucleus of about
0.06--0.12 km.
Using a central value of $D = 0.1$~km and the fast-rotating model
above with $L = 204$ kJ mol$^{-1}$, we found values of
$\dot{M}$ for the gas that fall within the uncertainty limits of
the \citet{Ci10} dust mass loss rate.

The agreement between the above model and the \citet{Ci10} observation
represents additional support for ${\cal M} \approx 1$ for inner
heliospheric mixtures of dust and gas in the vicinity of sublimating
bodies.
However, if we had chosen other values of ${\cal M}$, it would have
only required a small change in the latent heat to reproduce the
\citet{Je13} mass loss rate.
Specifically, the standard value of 
$L = 204$ kJ mol$^{-1}$ (which was optimized for ${\cal M}=1$)
needs to be decreased only to 172 kJ mol$^{-1}$ for ${\cal M}=0.01$,
or increased to 237 kJ mol$^{-1}$ for ${\cal M}=100$.
These changes in $L$ have been incorporated into an approximate
fitting formula for the sublimation rate $Z$ as a function of both
heliocentric distance and the dust-to-gas mass ratio:
\begin{equation}
  Z_{\rm fit} (r_{a}, {\cal M}) \, = \,
  Z_{0} \, {\cal M}^{-0.0427} \, r_{a}^{-2.25} \,
  \left[ 1 + ( r_{a} / \sigma )^{3} \right]^{-6} \,\, ,
\end{equation}
where $r_a$ is expressed in AU,
$Z_{0} = 2.545 \times 10^{16}$ particles cm$^{-2}$ s$^{-1}$,
and $\sigma = 0.0787 {\cal M}^{-0.0661}$.
This fitting formula is used below in order to vary ${\cal M}$
as a free parameter while retaining the empirical calibration to
Phaethon's dust mass loss rate.

\section{Coma and Tail Formation}
\label{sec:tail}

The goal of this section is to estimate a representative length scale 
for the dust-filled coma or tail surrounding an active asteroid
in the inner heliosphere.
This length scale is defined specifically as the largest distance from
the asteroid at which WISPR on {\em SPP} is expected to see
a dust-scattered enhancement over the sky background.
As above, we continue to use the observations of Phaethon at its perihelion
\citep{Je13} to help constrain some of the unknown parameters of the model.

\subsection{Spherically Symmetric Model}

\citet{Hu70} and \citet{DM71} computed the maximal radial distance
traversed by dust grains that are ejected from a comet's surface.
Newly freed grains become exposed immediately to sunlight
and begin to sublimate in the same manner as the parent asteroid.
A spherical dust grain with radius $a$ should have a finite lifetime
$\tau$ given by
\begin{equation}
  \tau \, = \, \frac{\rho a}{\mu m_{u} Z} \,\,\, ,
  \label{eq:tau}
\end{equation}
where $\rho$ is the mean mass density of escaping material.
In the models below, we assume a representative value
$\rho = 3$ g~cm$^{-3}$.
A dust grain that drifts away from the parent body with velocity
$v_{\rm dust}$ will thus traverse a radial distance of order 
$R \approx v_{\rm dust} \tau$ before it sublimates away completely.
However, \citet{Je13} speculated that the {\em observed} radial
distance of coma-like emission is probably going to be smaller than
$R$, because instrumental effects
(i.e., a high sky background and low photon counting statistics)
can obscure the faint outer parts of the dust cloud.
Thus, we need to estimate the visible-light brightness of a dust coma
(as a function of impact-parameter distance $b$ away from the asteroid)
and compare it to the sky background flux $F_Z$ defined in
Section \ref{sec:props}.

In lieu of a full three-dimensional model of the dynamics of dust
grains leaving the asteroid, an approximate spatial distribution
can be computed using the model of \citet{Ha57}.
Making use of mass flux conservation in spherical symmetry, with
a sink term to account for grain loss via sublimation, results in a
time-steady solution for the dust number density,
\begin{equation}
  n_{d}(r) \, = \, n_{d0} \left( \frac{R_a}{r} \right)^{2}
  \exp \left[ - \left( \frac{r - R_a}{v_{\rm dust} \tau}
  \right) \right] \,\, ,
  \label{eq:Haser}
\end{equation}
where $R_a$ is the asteroid radius and $r$ is the asteroid-centric
radial distance at which $n_d$ is measured.
The value of $n_{d0}$ at the asteroid surface can be estimated if
we know the dust-to-gas mass ratio ${\cal M}$ and the number
density of gas molecules $n_{g0}$ at the surface,
\begin{equation}
  n_{d0} \, = \, \frac{(\mu m_{u}) \, n_{g0}}{m_d} \,\, {\cal M}
\end{equation}
and $m_d$, the mass of an individual dust grain, can be computed
straightforwardly given its mean density $\rho$, radius $a$, and
the assumption it is roughly spherical in shape.
The Maxwellian gas mass loss theory (Equation \ref{eq:zmax})
also specifies
\begin{equation}
  n_{g0} \, = \, 4 Z \, / \, \bar{v}
\end{equation}
where the mean speed of gas molecules is given by
\begin{equation}
  \bar{v} \, = \, \sqrt{\frac{8 k_{\rm B} T}{\pi \mu m_{u}}}
  \label{eq:vgas}
\end{equation}
\citep[e.g.,][]{DM71}.

\begin{figure}
\hspace*{0.20in}
\includegraphics[width=0.85\textwidth]{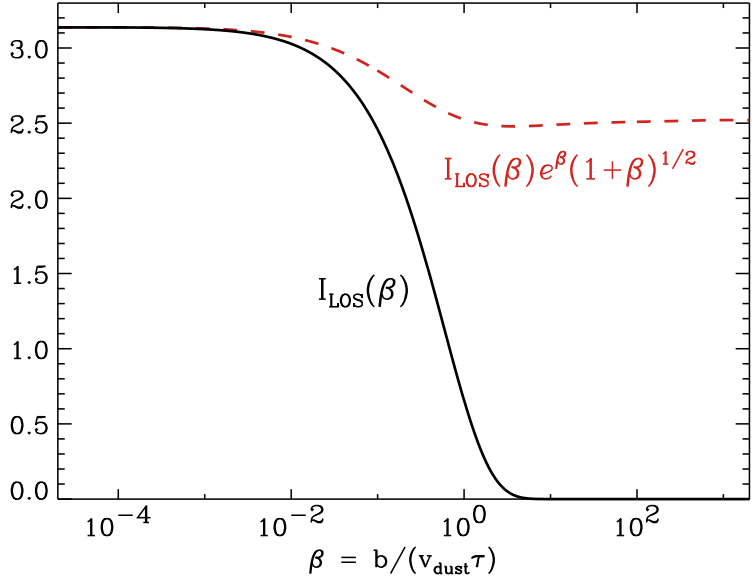}
\caption{Numerically integrated values of $I_{\rm LOS}$ versus
$\beta$ (solid black curve).  The dominant exponential behavior is
removed by multiplying $I_{\rm LOS}$ by the quantity
$e^{\beta} \sqrt{1 + \beta}$ (red dashed curve).}
\label{fig08}
\end{figure}

If a telescope views the dust emission at an impact-parameter distance
$b$ away from the asteroid itself, it will see dust grains with a
given column density $N(b)$.
To compute the column density, we assume the grains are distributed
with enough empty space around them so their observed cross sections
do not overlap.
Thus, $N(b)$ is given by an optically thin integral over the
line-of-sight distance $x$,
\begin{equation}
  N(b) \, = \, \int_{-\infty}^{+\infty} dx \,\,\, n_{d}(x)
  \, = \, \frac{n_{d0} \, R_{a}^2}{b}
  \, I_{\rm LOS} (\beta)
\end{equation}
where the dimensionless line-of-sight (LOS) integral is defined as
\begin{equation}
  I_{\rm LOS} (\beta) \, = \, \int_{-\infty}^{+\infty}
  \frac{du}{1 + u^2} \, \exp \left( - \beta \sqrt{1 + u^{2}} \right)
  \label{eq:ncol}
\end{equation}
and $\beta = b / (v_{\rm dust} \tau)$.
The above integral was derived by defining the heliocentric distance
$r$ of each point along the LOS using $r^{2} = x^{2} + b^{2}$,
and the dimensionless integration coordinate is $u = x/b$.
The above expression also assumes that $b \gg R_{a}$.
The integral was computed numerically over a fine grid in the
parameter $\beta$, and this was used as a lookup table when computing
actual column densities.
Figure~\ref{fig08} shows how $I_{\rm LOS}$ varies with $\beta$.

The visible-light flux $F_{\rm dust}$ emitted by dust grains (at a
given distance $b$ away from the asteroid, and in a solid angle
$\Omega$ that fills a WISPR pixel) can be computed and compared
to the background sky flux.
The total number of grains that contribute to this quantity is the
product of the column density $N(b)$ and the cross-sectional area of
one pixel as viewed by the spacecraft.
If the spacecraft--asteroid distance $d$ is assumed to be much larger
than the impact parameter $b$, the relevant cross-sectional area can
be estimated to be $\Omega d^{2}$.
Thus, if a single dust grain emits a visible-light flux $F_1$, the
total flux in the pixel is given by
\begin{equation}
  F_{\rm dust} \, = \, F_{1} \, N(b) \, \Omega d^{2}  \,\, .
  \label{eq:Fdust}
\end{equation}
The calculation of $F_1$ was done using the equations in
Section~\ref{sec:props} and the assumption that each dust grain is
essentially a ``tiny asteroid'' with a comparable albedo to its
parent body.

At any time during an encounter between an asteroid and {\em SPP,}
we can compute $F_{\rm dust}$ over a range of trial values of
the impact parameter $b$.
It is then possible to solve for a representative {\em observed coma
extent} ($b_{\rm coma}$) that is the distance at which $F_{\rm dust}$
is equal to a threshold sky background flux.
That sky flux was given by $10^{-4} F_{Z}$, which was determined
in Section~\ref{sec:props} to be the practical limit in modern-day
heliospheric imagers for resolving small features from a large-scale
background \citep{Df11}.
The quantity $F_Z$ depends on the elongation angle $\varepsilon$ and
the {\em SPP} heliocentric distance $r_p$.
Thus, the derived value of $b_{\rm coma}$ is observer-dependent and
not intrinsic to the asteroid.

In the limiting case of $\beta \ll 1$, the dimensionless integral
$I_{\rm LOS}$ approaches a constant value of $\pi$.
Thus, the solution for the observed coma size can be written as
\begin{equation}
  b_{\rm coma} \, \approx \, \left( \frac{F_1}{10^{-4} \, F_Z}
  \right) \, \pi R_{a}^{2} \, n_{d0} \, \Omega d^{2}
  \,\,\, .
\end{equation}
The above expression was used as a validation for the full numerical
solution in the limit of $b_{\rm coma} \ll v_{\rm dust} \tau$.

The model described above has three parameters that still have not
yet been specified: $\cal M$, $v_{\rm dust}$, and $a$.
We will use the \citet{Je13} measurement of
$b_{\rm coma} = 2.5 \times 10^{5}$~km for Phaethon at its perihelion
in order to constrain them.
Plausible ranges of variability for these parameters are given as follows:
\begin{enumerate}
\item
The dust-to-gas mass ratio ${\cal M}$ has been discussed above.
Comet observations appear to give values of ${\cal M} \sim 1$
\citep{AH95,Sa96,Ko07}, and recent high-quality measurements of
comet 67P/Churyumov-Gerasimenko indicate values between 3 and 10
\citep{Ro15,Fu16}.
Measurements for active asteroids do not yet exist.
Thus, with no firm firm observational guidance, we will vary ${\cal M}$
widely over seven orders of magnitude, between values of $10^{-3}$
and $10^{4}$.
\item
The grain velocity $v_{\rm dust}$ can be parameterized as a ratio
$\xi = v_{\rm dust}/v_{\rm gas}$,
since $v_{\rm gas}$ is known from the sublimation model.
\citet{DM71} found that escaping gas molecules eventually accelerate
to an asymptotic speed of $v_{\rm gas} \approx 1.8 \, \bar{v}$.
Thus, we specify $v_{\rm dust}$ as a function of $\bar{v}$ and $\xi$.
For asteroids in the inner heliosphere, the temperatures shown
in Figure~\ref{fig05}(b) indicate
$v_{\rm gas} \approx 0.3$--1 km~s$^{-1}$.
Comet observations tend to show that $v_{\rm gas}$ is a practical
upper limit for $v_{\rm dust}$
\citep[e.g.,][]{Wa92,Hu00,Pr04,Bee6,Bo08,Je12,Ig16}.
However, $v_{\rm dust}$ can take on a range of smaller values as well.
A lower limit on $v_{\rm dust}$ is the escape velocity
$v_{\rm esc} = (2GM_{a}/R_{a})^{1/2}$, which is typically between
$10^{-4}$ and $10^{-3}$ km~s$^{-1}$ for the asteroids considered here.
Thus, we will assume a plausible range for the dimensionless velocity
ratio of $10^{-4} < \xi < 1$.
\item
Astrophysical dust tends to exhibit a broad distribution of grain radii
$a$ rather than any one specific value \citep{Co94,Fu04}.
Observations of comets and the zodiacal light indicate particle sizes
spanning the range from 0.1~$\mu$m to 10~cm \citep{Fu93,Mn04,Hz06,KL15}.
Some models of dust loss from comets show a distinct anticorrelation
between $a$ and $v_{\rm dust}$ \citep[e.g.,][]{FP68,DM71,Ws82}, and
this can be used to estimate the value of $a$ that maximizes the
drift distance $R \approx v_{\rm dust} \tau$.
However, observations often show grain sizes in excess of this putative
maximum value \citep{Te11}.
These models also underestimate $v_{\rm dust}$ by giving values only
slightly larger than $v_{\rm esc}$, whereas observations often show
a range of higher speeds extending up to $v_{\rm gas}$ (see above).
In order to avoid undue dependence on these kinds of models, we allow
the value of $a$ to vary freely between 0.01~$\mu$m and 10~cm.
\end{enumerate}

With the above parameters ($\cal M$, $v_{\rm dust}$, $a$) varied freely
and the other asteroid properties fixed for Phaethon at its perihelion,
it is straightforward to compute $b_{\rm coma}$ using
Equations (\ref{eq:tau})--(\ref{eq:Fdust}) and compare the results
to the observed value of $2.5 \times 10^{5}$~km.
An initial search of the parameter space yielded several clear
constraints on the parameters.
To match the observed coma size, models with $\xi \leq 1$ must have
a dust-to-mass ratio of ${\cal M} > 0.48$ and a representative
grain size of $a < 39$~$\mu$m.
This latter constraint is in agreement with \citet{Je13}, who inferred
a grain radius of $a \approx 1$~$\mu$m for the particles producing the
emission in the observed coma of Phaethon.

\begin{figure}
\hspace*{0.20in}
\includegraphics[width=0.89\textwidth]{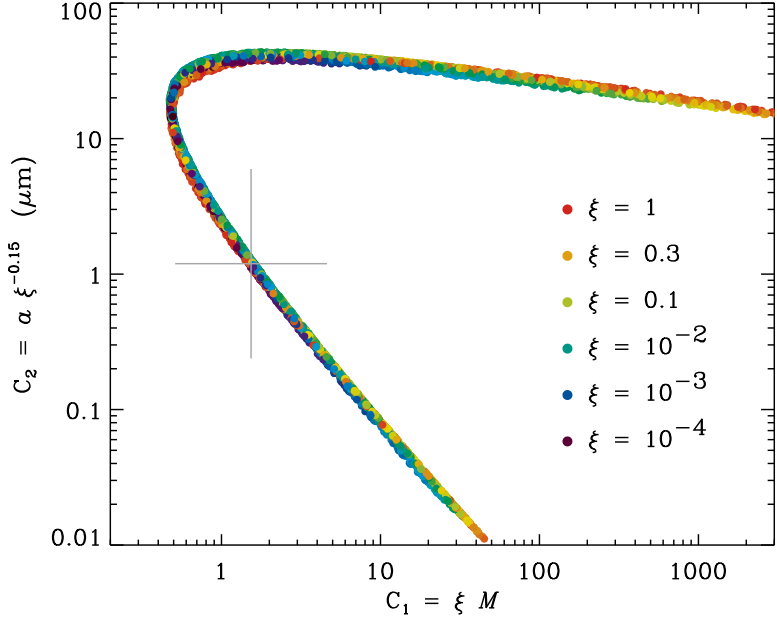}
\caption{Two-dimensional parameter space of solutions to for
$b_{\rm coma}$ that agree with observations of Phaethon at its
perihelion.
Symbol colors correspond to the values of $\xi$ (see legend), and
the gray cross indicates values adopted for use with the
{\em SPP} asteroid data: ${\cal M} = 3$, $\xi = 0.53$,
$a = 1$~$\mu$m.}
\label{fig09}
\end{figure}

In order to help better constrain valid ranges of the free parameters,
we searched for degeneracies (i.e., combinations of parameters that
produced identical values of $b_{\rm coma}$).
Figure~\ref{fig09} shows there is an extremely narrow ``allowed''
region in a two-dimensional cut through the solution space when the
orthogonal axis parameters are defined as
\begin{equation}
  C_{1} \, = \, \xi \, {\cal M}
  \,\,\,\, ,
  \,\,\,\,\,\,\,\,
  C_{2} \, = \, a \, \xi^{-0.15}  \,\,\, .
\end{equation}
The points shown in Figure~\ref{fig09} are a subset of results from a
Monte Carlo simulation of $10^6$ random trial solutions.
The 5363 displayed points represent only those solutions that agree
with the observed value of $b_{\rm coma}$ to within 1\%.

Because all of the points shown in Figure \ref{fig09} reproduce the
SECCHI observations of $b_{\rm coma}$ for Phaethon at perihelion,
we proceed by choosing one arbitrary set of values from those points
to use in the models below.
We presume the results for other asteroids at other heliocentric
distances should scale similarly no matter the details of this choice.
We follow \citet{Je13} by adopting $a = 1$~$\mu$m as a mean dust grain
radius.
With that constraint, the other two parameters are seen to follow the
relationship $\xi \approx 1.76 / {\cal M}^{1.085}$.
Optimized values of ${\cal M} = 3$ and $\xi = 0.53$ were thus chosen
in order to maintain continuity with the cometary analogy; i.e., the
empirical knowledge that ${\cal M}$ tends to be between 1 and 10 and
that $v_{\rm dust} \approx v_{\rm gas}$ (or at least 
$v_{\rm dust} \gg v_{\rm esc}$)
is often seen \citep[e.g.,][]{Je12,Ig16}.

\subsection{Results for SPP Asteroid Encounters}

\begin{figure}
\hspace*{0.20in}
\includegraphics[width=0.89\textwidth]{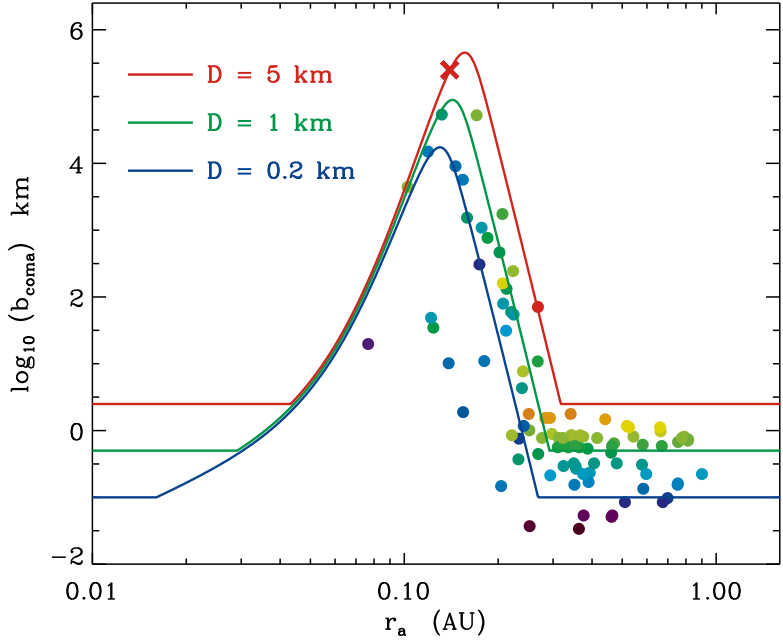}
\caption{Distance dependence of the modeled observable dust cloud size
$b_{\rm coma}$.
Symbols correspond to the 97 closest-approach events, with colors
corresponding to $H$ (i.e., asteroid diameter $D$) with the same
scalings as in Figure~\ref{fig02}.
Curves correspond to grids of models with idealized geometrical
configurations and fixed asteroid diameters (see text).
The measured tail length of Phaethon at its perihelion \citep{Je13}
is shown with a red X symbol.}
\label{fig10}
\end{figure}

Figure~\ref{fig10} shows some representative calculations of
$b_{\rm coma}$ for the set of 97 closest-approach events
described above (filled circles), and for a set of idealized asteroid
properties and positions (solid curves).
For these idealized curves, it was assumed that the three relevant
bodies (the asteroid, {\em SPP,} and the Sun) were situated on the
corners of an equilateral triangle.
In other words, for each point along these curves,
$r_{a} = r_{p} = d$ and $\alpha = \varepsilon = 60^{\circ}$.
For the 97 closest-approach events, these geometrical properties
were extracted from the ephemerides discussed earlier.
In cases when the dust ejected by an asteroid does not ever emit enough
flux to exceed the observable sky background, we set $b_{\rm coma}$ to
a lower limit of the asteroid radius $R_a$.
This happens for all modeled cases at $r_{a} > 0.3$~AU, which helps to
justify our choice for disregarding asteroids with perihelia $q > 0.3$~AU.

Note that there is a relatively finite range of heliocentric distances 
inside of which an asteroid is expected to emit grains that survive
for thousands of kilometers and produce a bright dust coma.
This range appears to be roughly $0.08 < r_{a} < 0.25$~AU.
For asteroids closer to the Sun than about 0.05~AU, the sublimation
rate $Z$ is so high that the grain lifetime $\tau$ is extremely short.
Thus, the grains cannot reach large values of $b$ before they are
destroyed.
For asteroids further away from the Sun than about 0.3~AU, the
sublimation rate $Z$ drops off to exceedingly small values.
This allows any escaping grains to essentially ``live forever,'' but
their number density $n_{d}$ is too low for their flux to compete
effectively with the sky background.

Rather than limit the calculation to the small database of 97
closest-approach events, the entire ephemeris for each pairing of
{\em SPP} with a given asteroid was processed to compute $b_{\rm coma}$
as a detailed function of time (i.e., at 0.1 day intervals).
To compare directly with planned observations with WISPR, each value of
$b_{\rm coma}$ was converted into a sky angle $\theta_{\rm coma}$
as observed from the vantage point of {\em SPP,}
\begin{equation}
  \theta_{\rm coma} \, = \, \tan^{-1} \left( 
  b_{\rm coma} / d \right)  \,\,\, .
\end{equation}
Because $b_{\rm coma}$ is a radius and not a diameter, a spherical
dust cloud should have an observable angular extent of
$2 \theta_{\rm coma}$.
However, we provide $\theta_{\rm coma}$ as a conservative lower
limit in cases of efficient tail ``blowback'' behind the asteroid.
A dust cloud is thus considered to be resolvable only when
$\theta_{\rm coma}$ exceeds the size of a single WISPR pixel
($\theta \approx 1.45'$).

\begin{figure}
\hspace*{0.20in}
\includegraphics[width=0.89\textwidth]{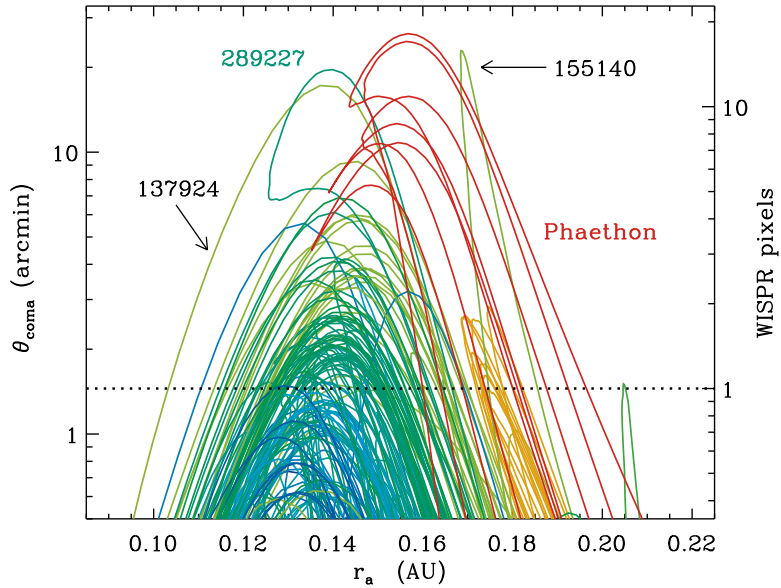}
\caption{Evolution of the modeled dust-cloud angular size
$\theta_{\rm coma}$ versus asteroid heliocentric distance for the
full set of modeled encounters with {\em SPP}.
Curve colors correspond to $H$ (i.e., asteroid diameter $D$) with the
same scalings as earlier figures.
The fiducial WISPR pixel size of 1.45$'$ is noted with a dotted line.}
\label{fig11}
\end{figure}

Figure~\ref{fig11} shows the full set of cases with modeled values
of $\theta_{\rm coma}$ that exceed 0.5$'$ on the sky.
The local maxima in $\theta_{\rm coma}$ occur neither at the times
when $d = d_{\rm min}$ nor at the times when $r_{a} = q$.
The large asteroid Phaethon shows up prominently with repeated peak
values of $\theta_{\rm coma}$ between 10$'$ and 26$'$, no matter the
location of {\em SPP} in the inner heliosphere.
The smaller asteroids 137924, 155140, and 289227 each have one-time
favorable events with $\theta_{\rm coma} > 17'$, but at other
perihelion passes the angular extents are smaller.
Figure~\ref{fig11} also makes clear that dust tail sizes greater
than about 1$'$ only tend to occur when asteroids fall between about
0.1 and 0.2 AU.
Together with the results shown in Figure~\ref{fig10}, this supports our
choice to disregard asteroids with perihelia greater than 0.2--0.3 AU.

\begin{figure}
\hspace*{0.001in}
\includegraphics[width=0.999\textwidth]{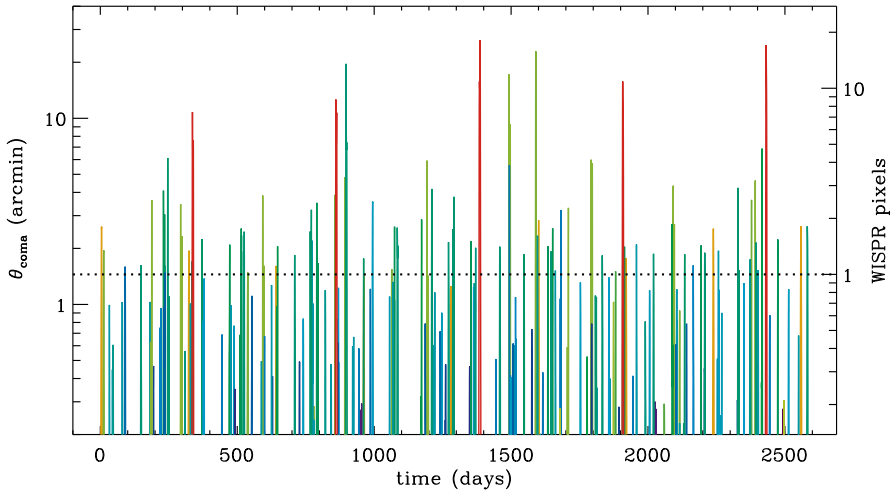}
\caption{Evolution of the modeled dust-cloud angular size
$\theta_{\rm coma}$ for various asteroids, versus mission time $t$ in
days.  Curve colors correspond to $H$ (i.e., asteroid diameter $D$)
with the same scalings as earlier figures.
The fiducial WISPR pixel size of 1.45$'$ is noted with a dotted line.}
\label{fig12}
\end{figure}

Figure~\ref{fig12} shows essentially the same information that is in
Figure~\ref{fig11}, but plotted as a function of mission time $t$
instead of heliocentric distance.
This shows that the individual episodes of large observable angular
extent are limited in time and distributed sporadically through the
{\em SPP} mission.
There are 113 predicted maxima that exceed the assumed WISPR pixel size,
corresponding to 24 distinct asteroids.
Several of the maxima are closely spaced pairs or triplets, separated
by hours to days, but most are isolated in time.
The mean duration of an event (defined as the time spent with
$\theta_{\rm coma} \geq 1.45'$) is 1.39 days, but the events cover a
range from 0.2 to 5.8 days.
Table~3 lists the most promising 41 of these events (i.e., only those
with $\theta_{\rm coma} \geq 3'$) in order of mission time, and also
gives the apparent magnitude $m_V$ of the parent asteroid and its
heliocentric distance at the specific times of maximum
$\theta_{\rm coma}$.

\begin{table}
\renewcommand\thetable{3}
\caption{Predictions for large angular dust tail events,
sorted by {\em SPP} mission time.}
\label{tab02}
\begin{tabular}{rlrcc}
\hline\noalign{\smallskip}
Time [days] &
Name &
$\theta_{\rm coma}$ [arcmin] &
$m_V$ &
$r_{a}$ [AU] \\
\noalign{\smallskip}\hline\noalign{\smallskip}
  187.90 &     394130   &   3.642 &  14.8 &  0.14501 \\
  230.71 &     394392   &   4.089 &  14.8 &  0.13994 \\
  234.80 &     394392   &   3.044 &  15.1 &  0.13974 \\
  246.21 &     374158   &   6.120 &  13.6 &  0.14030 \\
  293.41 &     137924   &   3.456 &  15.6 &  0.14530 \\
  336.51 &    Phaethon  &  10.813 &  12.1 &  0.15449 \\
  339.61 &    Phaethon  &   7.634 &  13.8 &  0.14855 \\
  593.10 &     137924   &   3.851 &  15.2 &  0.14837 \\
  770.76 &     374158   &   3.226 &  15.2 &  0.14133 \\
  790.45 &     394392   &   3.515 &  14.9 &  0.13631 \\
  857.14 &    2008 HW1  &   3.880 &  14.9 &  0.14726 \\
  860.34 &    Phaethon  &  12.624 &  11.8 &  0.15416 \\
  863.24 &    Phaethon  &  10.725 &  13.1 &  0.15071 \\
  893.13 &     137924   &   4.837 &  11.1 &  0.13828 \\
  897.13 &     289227   &  19.625 &  11.2 &  0.13989 \\
  897.23 &     137924   &   3.427 &  10.4 &  0.13533 \\
  899.03 &     289227   &   6.825 &  12.8 &  0.12748 \\
  899.83 &     289227   &   7.432 &  13.4 &  0.13680 \\
  994.40 &    2008 MG1  &   3.569 &  14.3 &  0.14497 \\
  994.60 &    2008 MG1  &   3.549 &  14.2 &  0.14520 \\
 1192.76 &     137924   &   5.927 &  13.1 &  0.14399 \\
 1210.85 &    2006 TC   &   4.175 &  14.7 &  0.14249 \\
 1291.13 &     374158   &   3.787 &  14.8 &  0.14196 \\
 1384.41 &    Phaethon  &  15.789 &  10.5 &  0.15016 \\
 1385.01 &    Phaethon  &  14.864 &  10.6 &  0.14506 \\
 1386.91 &    Phaethon  &  26.335 &  10.9 &  0.15667 \\
 1492.68 &     137924   &  17.238 &   9.6 &  0.13711 \\
 1493.88 &    2005 HC4  &   5.596 &  13.5 &  0.13366 \\
 1496.98 &     137924   &   9.264 &  12.2 &  0.14556 \\
 1590.26 &     155140   &  22.926 &   9.0 &  0.16850 \\
 1682.44 &   2013 HK11  &   3.211 &  12.5 &  0.15668 \\
 1709.33 &     394130   &   3.302 &  13.6 &  0.14533 \\
 1792.21 &     137924   &   5.994 &  14.2 &  0.14426 \\
 1796.71 &     137924   &   5.737 &  14.3 &  0.14657 \\
 1907.58 &    Phaethon  &  15.775 &  12.1 &  0.15692 \\
 2091.79 &     137924   &   4.344 &  15.0 &  0.14860 \\
 2328.12 &     374158   &   4.232 &  13.7 &  0.14082 \\
 2377.97 &    2008 HW1  &   3.645 &  14.4 &  0.14451 \\
 2391.48 &     137924   &   4.637 &  14.7 &  0.14859 \\
 2416.71 &     431760   &   6.875 &  13.1 &  0.14104 \\
 2431.22 &    Phaethon  &  24.707 &  11.1 &  0.15633 \\
\noalign{\smallskip}\hline
\end{tabular}
\end{table}

\section{Discussion and Conclusions}
\label{sec:conc}

The goal of this paper is to call the community's attention to the
likelihood that {\em SPP} will be well-positioned to observe mass loss
from Mercury-crossing asteroids in the inner heliosphere.
Specifically, we predict that there will be several times during
the {\em SPP} mission when the WISPR instrument will be able to
detect visible-light emission from the asteroids themselves and
(in a few cases) from associated dust clouds that may subtend almost
a degree of angular width on the sky.
These observations could fill in a large gap between the properties
of two heretofore distinct populations---active asteroids and
sungrazing comets---and thus help complete the census of primordial
solar system material.

Because of several ongoing uncertainties, many of the quantitative
predictions made above may not remain valid for the actual
{\em SPP} mission to commence in 2018.
Specifically, the following four factors will need to be re-evaluated
over the coming few years:
\begin{enumerate}
\item
As noted in Section \ref{sec:orb}, if the spacecraft launch 
slips from its nominal date of July 31, 2018, the
predicted time-dependent distances between {\em SPP} and the
asteroids will be incorrect.
The mission depends on multiple close encounters with
Venus to adjust the trajectory into the desired elliptical orbit
with a perihelion of 0.0459 AU.
\citet{Fx15} described a backup plan that involves a May 2019
launch plus one additional Venus gravity assist to bring {\em SPP}
into an orbit similar to the baseline trajectory.
\item
Many of the objects in our database of 97 Mercury-crossing asteroids
have been discovered relatively recently.
Thus, there may still be substantial uncertainties in their
ephemeris parameters.
Such errors would necessarily propagate into our predictions of the
relative times and distances of encounters with {\em SPP.}
Whether the improvement of these parameters would give rise to
a larger or smaller number of favorable encounters remains to be seen.
Nevertheless, the pace of asteroid discovery is likely to continue,
and there may be dozens more possible targets discovered between
now and 2018.
\item
The predictions made above did not take into account that the WISPR
instrument has a finite field of view and cannot see the entire sky.
Some fraction of favorable encounters with asteroids may end up
being hidden behind the {\em SPP} heat shield or other parts of
the spacecraft.
Thus, not every asteroid in listed Tables 1 and 3 will be observable at
all times.
These details need to be considered when constructing detailed
observation plans, but they are beyond the scope of this paper.
\item
The observability of any given asteroid and its surrounding dust cloud
was computed using a threshold sky background of $10^{-4} F_Z$
\citep[see, e.g.,][]{Df11}.
It is possible that WISPR may contain sufficient improvements
in photon counting or flat-fielding, relative to the SECCHI package
on {\em STEREO,} that could allow even weaker signals to be extracted
from the raw images.
\end{enumerate}
However the above issues are resolved, there appears to be a high
probability for a significant number of encounters between {\em SPP}
and Mercury-crossing asteroids during times when the latter may
be losing mass at $r_{a} < 0.2$~AU.
Details aside, the statistical distribution of events is likely to
remain similar to what was computed in this paper.

In addition to refining the positional and temporal accuracy of
the above predictions, there are also several ways
that the mass loss modeling can be improved.
Our assumption of constant values for $L$ and $\mu$ should be
replaced by a more self-consistent (i.e., temperature dependent)
description of specific ejected materials.
This is particularly important because the computed rates
are often in the exponentially dropping part of the sublimation
curve (Figure \ref{fig05}), where small variations in the input
parameters could change $Z$ by several orders of magnitude.
Also, the spherically symmetric \citet{Ha57} model should be replaced
by a full three-dimensional dynamical simulation of the ejected
dust grains.
If the escaping grains are swept back into a collimated tail,
their number density may be up to an order of magnitude higher (and
thus more easily observable when viewed from a favorable direction)
than if they were spread out in a spherical cloud.
\citet{Je11} estimated the sunward turnaround distance $s$ expected from
grains of a given size.
In cases where $s \ll b_{\rm coma}$ it would be most useful to apply
such a three-dimensional correction to the density model.
At the very least, the standard \citet{FP68} type of ballistic
modeling should be done for the most promising encounters, in order
to predict the tail orientations and geometries.

The idea of using inner heliospheric space probes as remote observatories
for active asteroids should be expanded beyond just {\em SPP.}
The {\em Solar Orbiter} mission \citep{Mu13,Be15} will reach a minimum
perihelion distance of about 0.28 AU during a similar time-frame
as {\em SPP,} but it will also leave the ecliptic plane to eventually
reach inclination angles of order 30$^{\circ}$.
\citet{Sa15} proposed a mission to visit Phaethon and
associated asteroids (155140) 2005 UD and (225416) 1999 YC, which
would explore the origins of the Geminid meteor stream and study the
physics of comet/asteroid transition objects.
Near-Earth asteroid flyby or rendezvous missions with infrared
spectro-imagers \citep{Gr16} could also improve our knowledge
of the physics of regolith loss in a hot thermal environments.
Lastly, any spacecraft that comes close enough to fly through
the dust tail of an active asteroid would put unprecedented
constraints on the properties of the ejected grains, thus allowing
a comparison of similarities and differences to dust ejected by comets
\citep[see, e.g.,][]{KK87,MC05,Ne07,DC15}.

\begin{acknowledgements}
The author gratefully acknowledges Kelly Korreck and Martha Kusterer
for supplying the {\em SPP} SPICE kernel.
The initial idea for this paper grew out of online discussions with
Corey Powell and Matthew Francis about interesting new destinations
for space probes after the July 2015 {\em New Horizons} flyby of Pluto.
The author also thanks David Malaspina for helpful comments on the
manuscript, and the anonymous referees for many constructive suggestions.
This work was supported by start-up funds from the Department of
Astrophysical and Planetary Sciences at the University of Colorado
Boulder.  This research made extensive use of NASA's Astrophysics
Data System.
\end{acknowledgements}

\end{document}